\documentclass[fleqn,usenatbib]{mnras}

\usepackage{newtxtext,newtxmath}

\usepackage[T1]{fontenc}
\usepackage{ae,aecompl}

\usepackage{graphicx}	
\usepackage{amsmath}	
\usepackage{amssymb}	
\usepackage{tabularx}

\newcommand{\msun}{\ensuremath{\mbox{M}_{\odot}}}
\newcommand{\lsun}{\ensuremath{\mbox{L}_{\odot}}}
\newcommand{\rsun}{\ensuremath{\mbox{R}_{\odot}}}

\title[The rotation of $\alpha$~Oph]{The rotation of $\alpha$~Oph investigated using polarimetry }

\author[J. Bailey et al.]{Jeremy Bailey$^{1}$\thanks{E-mail: j.bailey@unsw.edu.au},
Daniel V. Cotton$^{2,3,4}$,
Ian D. Howarth$^{5}$,
Fiona Lewis$^{1}$,
\newauthor Lucyna Kedziora-Chudczer$^{4}$
\\
\\
$^{1}$School of Physics, University of New South Wales, Sydney, NSW 2052, Australia.\\
$^{2}$Anglo Australian Telescope, Australian National University, 418 Observatory Road, Coonabarabran, NSW 2357, Australia.\\
$^{3}$Western Sydney University, Locked Bag 1797, Penrith-South DC, NSW 1797, Australia.\\
$^{4}$Centre for Astrophysics, University of Southern Queensland, Toowoomba, Queensland 4350, Australia.\\
$^{5}$University College London, Gower Street, London WC1E 6BT, UK.\\
}

\date{Accepted XXX. Received YYY; in original form ZZZ}

\pubyear{2020}

\begin{document}
\label{firstpage}
\pagerange{\pageref{firstpage}--\pageref{lastpage}}
\maketitle

\begin{abstract}
Recently we have demonstrated that high-precision polarization observations can detect the polarization resulting from the rotational distortion of a rapidly rotating B-type star. Here we investigate the extension of this approach to an A-type star. Linear-polarization observations of $\alpha$~Oph (A5IV) have been obtained over wavelengths from 400 to 750 nm. They show the wavelength dependence expected for a rapidly-rotating star combined with a contribution from interstellar polarization. We  model the observations by fitting rotating-star polarization models and adding additional constraints including a measured $v_{\rm e}\sin{i}$. However, we cannot fully separate the effects of rotation rate and inclination, leaving a range of possible solutions. We determine a rotation rate ($\omega = \Omega/\Omega_{\rm c}$) between 0.83 and 0.98 and an axial inclination $i > 60^\circ$. The rotation-axis position angle is found to be $142^\circ \pm 4^\circ$,  differing by 16$^\circ$ from a value obtained by interferometry. This might be due to precession of the rotation axis due to interaction with the binary companion. Other parameters resulting from the analysis include a polar temperature  $T_{\rm p} = 8725 \pm 175$~K, polar gravity  $\log{g_{\rm p}} = 3.93 \pm 0.08$ (dex cgs),  and polar radius $R_{\rm p} = 2.52 \pm 0.06~$\rsun. Comparison with rotating-star evolutionary models indicates that $\alpha$~Oph is in the later half of its main-sequence evolution and must have had an initial $\omega$ of 0.8 or greater. The interstellar polarization has a maximum value at a wavelength ($\lambda_{\rm max}$) of $440 \pm 110$~nm, consistent with values found for other nearby stars.
\end{abstract}




\section{Introduction}

Hot stars can produce polarized light through scattering by electrons in their atmospheres, as first suggested by \citet{chandrasekhar46}. Integrated over a spherical star the polarization will average to zero, but if there is a departure from spherical symmetry a net polarization can be observed. \citet{harrington68} suggested that one way of producing the required asymmetry would be the rotational distortion of a rapidly-rotating star. However, subsequent studies using more complete stellar-atmosphere models showed that the expected polarization was small at visible wavelengths \citep{collins70,sonneborn82}. It was not until the development of polarimeters that could measure linear polarization to parts-per-million (ppm) levels of precision \citep{hough06,bailey15} that the effect was detected \citep{bailey10} and confirmed by multiwavelength polarimetry and detailed modelling of the bright star Regulus ($\alpha$~Leo; \citealt{cotton17}).

In the case of Regulus (B7IV) the polarimetric observations showed a distinctive wavelength dependence  with the polarization direction changing from parallel to the star's rotation axis at red wavelengths to perpendicular to the rotation axis at blue wavelengths \citep{cotton17}, as predicted by models \citep{sonneborn82}. Detailed modelling of this pattern then allowed a determination of the rotation rate ($\omega = 0.965^{+0.006}_{=0.008}$), inclination (> 76.5$\degr$), and rotation-axis position angle (79.5 $\pm$ 0.7$\degr$; \citealt{cotton17}) that were in good agreement with independent determinations from interferometric imaging \citep{che11}.

In this paper we apply the same type of analysis to the A-type star $\alpha$~Oph (Rasalhague, HD~159561). This is a rapidly-rotating star of spectral type A5IV \citep{gray01} at a distance of 14.9 pc \citep{vanLeeuwen07}. It is in a binary system, detected as an astrometric binary \citep{lippincott66} and subsequently resolved using speckle interferometry and adaptive-optics imaging \citep{mccarthy83, hinkley11}. It is also a $\delta$-Scuti-type pulsating variable \citep{monnier10}.

\begin{figure}
	\includegraphics[trim={0cm 4cm 0cm 0cm}, width=\columnwidth]{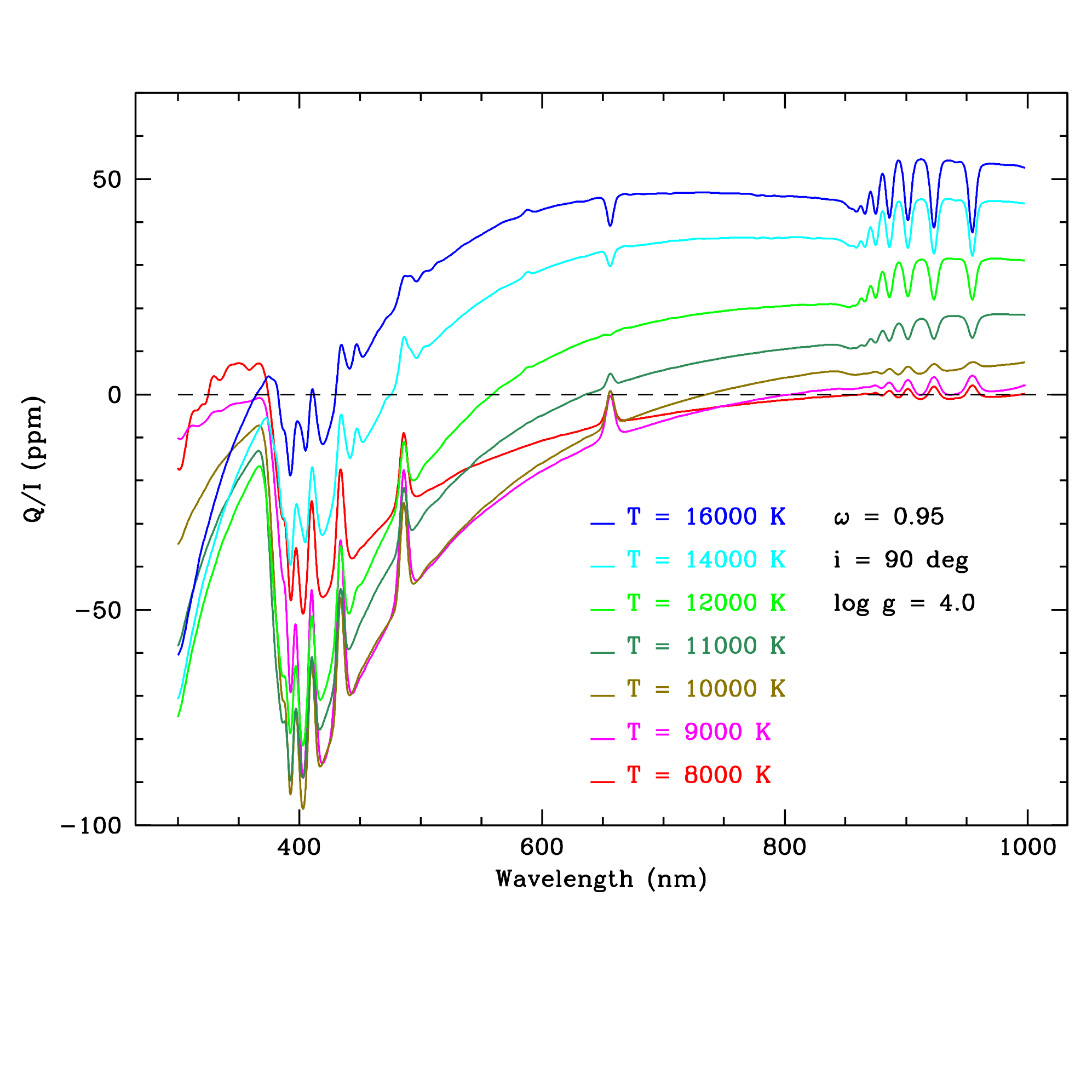}
    \caption{Modelled polarization for rapidly-rotating stars of different polar temperatures with $\omega$ = 0.95, inclination 90 degrees and $\log{g}$ = 4.0. Modelling uses the methods described in \citet{cotton17} and in section~\ref{sec_model} of this paper.}
    \label{fig_pol_95}
\end{figure}

Polarization is generally expected to be lower for cooler stars as the atmosphere is less ionized and there are fewer scattering electrons per unit mass. We therefore did some preliminary calculations to investigate the expected polarization levels in cooler rotating stars. In Fig.~\ref{fig_pol_95} we show the modelled polarization for rapidly rotating ($\omega$ = 0.95) stars of different polar temperatures. The models were calculated using the methods described in \citet{cotton17} and in section~\ref{sec_model} of this paper. The polarization wavelength dependence seen in these models is a complex combination of several effects. Polarization depends on the balance between scattering and absorption, so polarization generally drops on strong absorption lines and shortward of the Balmer and Paschen jumps where bound-free continuum absorption is strong. It also depends on the variation of polarization across the star due to changing viewing angle and changing temperatures due to gravity darkening \citep[e.g. Fig. 2 of][]{cotton17}. There is a distinctive change in sign of polarization in the hotter models as described by \citet{harrington17}.
It can be seen that significant polarization is seen at all the temperatures modelled. At the lower temperatures typical of A stars the models predict that polarization is seen mostly at blue-green wavelengths and is close to zero at the red end of the spectrum. The polarization is mostly negative, which means that it is perpendicular to the star's rotation axis. This modelling shows that significant polarization due to rotation should be observable in stars down to temperatures of $\sim$8000 K.

In this paper we report polarimetric observations of $\alpha$~Oph in section~\ref{sec_obs}. We compare the data with models using a similar approach to that adopted in \citet{cotton17} in section~\ref{sec_model}, and discuss the results in section~\ref{sec_discuss}.

\section{Observations}
\label{sec_obs}

\begin{table*}
\caption{Summary of Observing Runs and TP Measurements}
\label{tab:runs}
\tabcolsep 3.5 pt
\begin{tabular}{cl|ccrcccc|ccc|rr}
\hline
\hline
\multicolumn{2}{c|}{Run} & \multicolumn{7}{c|}{Telescope and Instrument Set-Up$^a$}   &   \multicolumn{3}{c|}{Observations$^b$}   &   \multicolumn{2}{c}{Calibration$^{cd}$}    \\
Desig. & \multicolumn{1}{c|}{Date Range$^e$} & Instr. &  Tel. & \multicolumn{1}{c}{f/} & Ap. & Mod. & Fil. & Det.$^f$ & n & $\lambda_{\rm eff}$ &  Eff. & \multicolumn{1}{c}{$q_{TP}$} & \multicolumn{1}{c}{$u_{TP}$} \\
 &  \multicolumn{1}{c|}{(UT)} &  &   &  & ($\arcsec$) &  &  &  &  & (nm) & ($\%$) & \multicolumn{1}{c}{(ppm)} & \multicolumn{1}{c}{(ppm)} \\
\hline
2005APR  & 2005-04-27     & PlanetPol  & WHT  &11\phantom{*} &  5.2 & PEM    & BRB         & APD & 1 & 753.8 & 93.0 & \multicolumn{2}{c}{\citet{bailey10}$^g$} \\
\hline
2016JUL  & 2016-09-19     & M-HIPPI    & UNSW & 11\phantom{*} & 58.9 & MT     & Clear       & B   & 1 & 477.9 & 82.6 & $-$70.1 $\pm$ 2.9 &$-$10.1 $\pm$ 2.9 \\
\hline
2017AUG  & 2017-08-07 to 18 & HIPPI    & AAT  & 8\phantom{*} &  6.6 & BNS-E2 & 425SP       & B   & 1 & 400.9 & 52.0 &  $-$7.3 $\pm$ 3.6 &  8.5 $\pm$ 3.6 \\
&                         &            &      &   &      &        & 500SP       & B   & 1 & 438.0 & 75.9 & $-$10.0 $\pm$ 1.7 & $-$0.4 $\pm$ 1.6 \\
&                         &            &      &   &      &        & g$^{\prime}$& B   & 1 & 466.3 & 87.3 &  $-$9.1 $\pm$ 1.5 & $-$2.6 $\pm$ 1.4 \\
&                         &            &      &   &      &        & r$^{\prime}$& R   & 1 & 620.8 & 82.4 & $-$10.4 $\pm$ 1.3 & $-$7.0 $\pm$ 1.3 \\
&                         &            &      &   &      &        & 650LP       & R   & 2 & 718.9 & 65.7 &  $-$8.2 $\pm$ 2.3 & $-$5.1 $\pm$ 2.4 \\
\hline
2018MAR$^h$& 2018-04-04   & HIPPI-2    & AAT  & 8*& 15.7 & BNS-E3 & g$^{\prime}$& B   & 0 & 467.4 & 82.6 & 130.1 $\pm$ 0.9 & 3.9 $\pm$ 0.9 \\
\hline
2018JUN & 2018-07-05 to 06  & HIPPI-2  & GN   &16\phantom{*} &  6.4 & ML-E1  & r$^{\prime}$& B   & 2 & 603.3 & 61.5 & \multicolumn{2}{c}{\citet{bailey19b}$^g$} \\
\hline
2018JUL & 2018-07-15 to 23  & HIPPI-2  & AAT  & 8*& 11.9 & BNS-E4 & 425SP       & B   & 2 & 403.1 & 38.3 &  $-$5.6 $\pm$ 6.4 & 19.8 $\pm$ 6.3 \\
&                         &            &      &   &      &        & 500SP       & B   & 2 & 440.1 & 68.2 &   1.9 $\pm$ 1.4 & 18.4 $\pm$ 1.4 \\
&                         &            &      &   &      &        & g$^{\prime}$& B   & 2 & 464.8 & 80.4 & $-$12.8 $\pm$ 1.1 &  4.1 $\pm$ 1.0 \\
&                         &            &      &   &      &        & V           & B   & 2 & 533.4 & 95.5 & $-$20.3 $\pm$ 1.5 &  2.3 $\pm$ 1.5 \\
&                         &            &      &   &      &        & r$^{\prime}$& B   & 2 & 602.7 & 86.7 & $-$10.4 $\pm$ 2.2 &  3.7 $\pm$ 2.2 \\
&                         &            &      &   &      &        & r$^{\prime}$& R   & 1 & 622.3 & 82.9 & $-$12.7 $\pm$ 1.2 &  0.4 $\pm$ 1.2 \\
&                         &            &      &   &      &        & 650LP       & R   & 1 & 722.3 & 64.8 &  $-$6.6 $\pm$ 1.9 &  4.0 $\pm$ 1.9 \\
\hline
2018AUG & 2018-08-18 to 22 & HIPPI-2   & AAT  & 8*& 11.9 & BNS-E5 & 425SP       & B   & 1 & 403.1 & 24.5 &  $-$5.6 $\pm$ 6.4 & 19.8 $\pm$ 6.3 \\
&                         &            &      &   &      &        & V           & B   & 1 & 535.5 & 95.2 & $-$20.3 $\pm$ 1.5 &  2.3 $\pm$ 1.5 \\
&                         &            &      &   &      &        & r$^{\prime}$& B   & 1 & 602.7 & 92.8 & $-$10.4 $\pm$ 2.2 &  3.7 $\pm$ 2.2 \\
\hline
2019MAR & 2019-03-26      & HIPPI-2    & AAT  &15\phantom{*} & 12.7 & ML-E1  & 425SP       & B   & 3 & 399.8 & 71.4 &  $-$1.7 $\pm$ 1.1 &  4.2 $\pm$ 1.0 \\
\hline
2019APR & 2019-04-20      & HIPPI-2    & AAT  &15\phantom{*} & 12.7 & ML-E1  & V           & B   & 1 & 533.0 & 82.2 &  10.4 $\pm$ 1.5 & 57.4 $\pm$ 1.5 \\
&                         &            &      &   &      &        & r$^{\prime}$& B   & 1 & 602.7 & 61.6 &   0.7 $\pm$ 2.3 & 14.1 $\pm$ 2.4 \\
\hline
\hline
\end{tabular}
\begin{flushleft}
Notes: \\
\textbf{*} Indicates use of a 2$\times$ negative achromatic lens, effectively making the foci f/16.\\ 
\textbf{$^a$} A full description along with transmission curves for all the components and modulation characterisation of each modulator in the specified performance era can be found in \citet{bailey19b}.\\ 
\textbf{$^b$} Median values are given as representative of the observations made of $\alpha$~Oph. Exact values are given in Table~\ref{tab:observations} for each observation.\\
\textbf{$^c$} The observations used to determine the telescope polarization (TP) for the Mini-HIPPI run are described in \citet{bailey17}, those for HIPPI-2 up until 2019MAR are described in \citet{bailey19b}; TPs for the 2019APR run are made up of, V: 3$\times$ HD~49815, 2$\times$ HD~102647; r$^{\prime}$: 3$\times$ HD~49815, 1$\times$ HD~102647; and for the 2017AUG HIPPI run, 425SP: 2$\times$ HD~2151, 3$\times$ HD~48915, 2$\times$ HD~102647, 2$\times$ HD~102870; 500SP: 2$\times$ HD~2151, 3$\times$ HD~48915, 2$\times$ HD~102647, 2$\times$ HD~102870; g$^{\prime}$: 2$\times$ HD~2151, 2$\times$ HD~48915, 2$\times$ HD~102647, 2$\times$ HD~102870; r$^{\prime}$: 3$\times$ HD~2151, 3$\times$ HD~48915, 2$\times$ HD~102647, 2$\times$ HD~102870; 650LP: 3$\times$ HD~2151, 3$\times$ HD~48915, 1$\times$ HD~102870.\\ 
\textbf{$^d$} The high-polarization standards observed to calibrate position angle, and the values of $\sigma_{PA}$ for the HIPPI-2 runs up until 2019MAR are given in \citet{bailey19b}, and those for the Mini-HIPPI run are given in \citet{bailey17}; for the 2019APR run the standards observed in g$^{\prime}$ were: HD~80558, 2$\times$ HD~147084, HD~111613, HD~187929 giving $\sigma_{PA}=0.27^{\circ}$, based on a smaller number of observations (one each of HD~147084 and HD~187929 in each band), corrections were made to different bands as follows: V -2.145$^{\circ}$, r$^{\prime}$ 1.575$^{\circ}$; for the 2017AUG HIPPI run the g$^{\prime}$ observations were: HD~147084, HD~154445, HD~187929 giving $\sigma_{PA}=0.53^{\circ}$.\\   
\textbf{$^e$} Dates given are for observations of $\alpha$~Oph and/or control stars. \\
\textbf{$^f$} APD means `avalanche photodiode'; B, R indicate blue- and red-sensitive H10720-210 and H10720-20 PMTs, respectively.\\
\textbf{$^g$} The TP functions for the AltAz telescopes are more complicated, and are given in the papers cited.\\
\textbf{$^h$} No observations of $\alpha$~Oph were made during the 2018MAR run, but a control star was observed. Figures quoted are for HD~182640. \\
\end{flushleft}
\label{tab:mod}
\end{table*}

\begin{table*}
\caption{Polarization observations of $\alpha$~Oph}
\label{tab:observations}
\tabcolsep 4 pt
\begin{tabular}{llrrccrrrrrr}
\hline
\hline
Run     & UT                  & Dwell & Exp. & Fil. & Det.$^a$ & $\lambda_{\rm eff}$ & Eff. & \multicolumn{1}{c}{$q$} & \multicolumn{1}{c}{$u$} & \multicolumn{1}{c}{$p$} & \multicolumn{1}{c}{$\theta$}\\
        &                       & \multicolumn{1}{c}{(s)} & \multicolumn{1}{c}{(s)} & & & \multicolumn{1}{c}{(nm)} & \multicolumn{1}{c}{(\%)} & \multicolumn{1}{c}{(ppm)} & \multicolumn{1}{c}{(ppm)} & \multicolumn{1}{c}{(ppm)} & \multicolumn{1}{c}{($^\circ$)}\\
\hline
2019MAR & 2019-03-26 18:49:01 & 752 & 480 & 425SP & B & 399.6 & 71.3 &  $-$19.3 $\pm$ 12.9 &   58.9 $\pm$ 12.8 & 62.0 $\pm$ 12.9 &   54.1 $\pm$ \phantom{0}6.0\\
2019MAR & 2019-03-26 18:11:02 & 936 & 640 & 425SP & B & 399.7 & 71.3 &   18.4 $\pm$ 12.4 &   77.6 $\pm$ 12.4 &   79.8 $\pm$ 12.4 &   38.3 $\pm$ \phantom{0}4.5\\
2019MAR & 2019-03-26 17:14:41 & 940 & 640 & 425SP & B & 400.0 & 71.6 &   16.5 $\pm$ 12.4 &   34.7 $\pm$ 12.5 &   38.4 $\pm$  12.5 &   32.3 $\pm$   \phantom{0}9.8\\
2017AUG & 2017-08-10 10:26:44 & 2807 & 1920 & 425SP & B & 400.9 & 52.0 &   $-$2.9 $\pm$ 14.2 &   18.8 $\pm$ 14.2 & 19.0 $\pm$  14.2 &   49.4 $\pm$  26.1\\
\hline
2018JUL & 2018-07-15 11:25:49 & 1438 & 960 & 425SP & B & 403.1 & 38.3 &   45.5 $\pm$ 14.6 &   74.1 $\pm$ 14.5 &   87.0 $\pm$  14.6 &   29.2 $\pm$   \phantom{0}4.9\\
2018JUL & 2018-07-19 11:11:13 & 1544 & 960 & 425SP & B & 403.1 & 38.3 &   56.5 $\pm$ 14.9 &   63.1 $\pm$ 14.8 &   84.7 $\pm$  14.8 &   24.1 $\pm$   \phantom{0}5.1\\
2018AUG & 2018-08-22 10:17:18 & 1315 & 960 & 425SP & B & 403.1 & 24.5 &   79.4 $\pm$ 17.1 &  141.4 $\pm$ 17.0 &  162.2 $\pm$  17.1 &   30.3 $\pm$   \phantom{0}3.0\\
\hline
2017AUG & 2017-08-10 11:17:04 & 1859 & 640 & 500SP & B & 438.0 & 75.9 &   16.6 $\pm$ \phantom{0}8.1 &   52.8 $\pm$ \phantom{0}8.0 &   55.3 $\pm$   \phantom{0}8.1 &   36.3 $\pm$   \phantom{0}4.2\\
2018JUL & 2018-07-15 11:48:51 & 1060 & 640 & 500SP & B & 440.0 & 68.1 &   21.0 $\pm$ \phantom{0}6.8 &   57.3 $\pm$ \phantom{0}6.8  &   61.0 $\pm$   \phantom{0}6.8 &   34.9 $\pm$   \phantom{0}3.2\\
2018JUL & 2018-07-19 10:47:45 & 1067 & 640 & 500SP & B & 440.1 & 68.2 &   22.8 $\pm$ \phantom{0}7.0 &   63.7 $\pm$ \phantom{0}7.0 &   67.7 $\pm$   \phantom{0}7.0 &   35.2 $\pm$   \phantom{0}3.0\\
\hline
2018JUL & 2018-07-15 12:07:14 & 1056 & 640 & g$^{\prime}$ & B & 464.8 & 80.4 &   14.9 $\pm$ \phantom{0}2.9 &   52.5 $\pm$ \phantom{0}2.8 &   54.6 $\pm$   \phantom{0}2.8 &   37.1 $\pm$   \phantom{0}1.5\\
2018JUL & 2018-07-19 11:35:38 & 1092 & 640 & g$^{\prime}$ & B & 464.8 & 80.4 &   20.2 $\pm$ \phantom{0}2.9 &   65.9 $\pm$ \phantom{0}2.9 &   68.9 $\pm$   \phantom{0}2.9 &   36.5 $\pm$   \phantom{0}1.2\\
2017AUG & 2017-08-10 11:16:18 & 2436 & 640 & g$^{\prime}$ & B & 466.3 & 87.3 &   14.5 $\pm$ \phantom{0}3.4 &   60.3 $\pm$ \phantom{0}3.5 &   62.0 $\pm$   \phantom{0}3.5 &   38.2 $\pm$   \phantom{0}1.6\\
\hline
2016JUL & 2016-09-19 10:40:34 & 1450 & 800 & Clear & B & 477.9 & 82.6 &    6.9 $\pm$ 23.2 &   65.2 $\pm$ 23.8 &   65.6 $\pm$  23.5 &   42.0 $\pm$  11.2\\
\hline
2019APR & 2019-04-20 16:11:51 & 846 & 480 & V & B & 533.0 & 82.2 &   36.6 $\pm$ \phantom{0}5.2 &   47.4 $\pm$ \phantom{0}5.0 &   59.9 $\pm$   \phantom{0}5.1 &   26.2 $\pm$   \phantom{0}2.5\\
2018JUL & 2018-07-15 12:26:01 & 1047 & 640 & V & B & 533.4 & 95.5 &   24.3 $\pm$ \phantom{0}3.8 &   39.7 $\pm$ \phantom{0}3.8 &   46.5 $\pm$   \phantom{0}3.8 &   29.3 $\pm$   \phantom{0}2.3\\
2018JUL & 2018-07-19 11:54:21 & 1058 & 640 & V & B & 533.4 & 95.5 &   34.2 $\pm$ \phantom{0}3.9 &   60.1 $\pm$ \phantom{0}3.8 &   69.1 $\pm$   \phantom{0}3.8 &   30.2 $\pm$   \phantom{0}1.6\\
2018AUG & 2018-08-22 10:38:16 & 983 & 640 & V & B & 533.5 & 95.2 &   31.7 $\pm$ \phantom{0}4.1 &   41.3 $\pm$ \phantom{0}4.1 &   52.1 $\pm$   \phantom{0}4.1 &   26.2 $\pm$   \phantom{0}2.2\\
\hline
2018JUL & 2018-07-15 12:44:35 & 1031 & 640 & r$^{\prime}$ & B & 602.6 & 86.7 &   21.0 $\pm$ \phantom{0}6.2 &   42.3 $\pm$ \phantom{0}6.2 &   47.2 $\pm$   \phantom{0}6.2 &   31.8 $\pm$   \phantom{0}3.8\\
2019APR & 2019-04-20 15:57:17 & 832 & 480 & r$^{\prime}$ & B & 602.7 & 61.6 &   21.0 $\pm$ \phantom{0}9.4 &   28.9 $\pm$ \phantom{0}9.5 &   35.7 $\pm$   \phantom{0}9.4 &   27.0 $\pm$   \phantom{0}7.7\\
2018JUL & 2018-07-19 10:28:50 & 1092 & 660 & r$^{\prime}$ & B & 602.7 & 86.7 &   20.0 $\pm$ \phantom{0}6.0 &   40.5 $\pm$ \phantom{0}6.3 &   45.2 $\pm$   \phantom{0}6.1 &   31.9 $\pm$   \phantom{0}3.8\\
2018AUG & 2018-08-22 10:57:47 & 1317 & 960 & r$^{\prime}$ & B & 602.7 & 92.8 &   31.2 $\pm$ \phantom{0}5.2 &   30.3 $\pm$ \phantom{0}4.9 &   43.5 $\pm$   \phantom{0}5.1 &   22.1 $\pm$   \phantom{0}3.3\\
2018JUN & 2018-07-05 08:22:32 & 573 & 192 & r$^{\prime}$ & B & 603.2 & 61.5 &   17.2 $\pm$ 11.0 &   48.8 $\pm$ 11.1 &   51.7 $\pm$  11.0 &   35.3 $\pm$   \phantom{0}6.2\\
2018JUN & 2018-07-06 11:59:18 & 488 & 160 & r$^{\prime}$ & B & 603.3 & 61.4 &    1.9 $\pm$ 10.7 &   40.5 $\pm$ 10.9 &   40.5 $\pm$  10.8 &   43.7 $\pm$   \phantom{0}7.8\\
\hline
2017AUG & 2017-08-07 11:41:29 & 2788 & 1920 & r$^{\prime}$ & R & 620.8 & 82.4 &   26.2 $\pm$ \phantom{0}2.9 &   30.9 $\pm$ \phantom{0}2.9 &   40.5 $\pm$   \phantom{0}2.9 &   24.9 $\pm$   \phantom{0}2.0\\
2018JUL & 2018-07-23 10:21:11 & 970 & 640 & r$^{\prime}$ & R & 622.3 & 82.9 &   27.7 $\pm$ \phantom{0}3.4 &   39.8 $\pm$ \phantom{0}3.4 &   48.5 $\pm$   \phantom{0}3.4 &   27.6 $\pm$   \phantom{0}2.0\\
\hline
2017AUG & 2017-08-07 10:25:45 & 2217 & 1280 & 650LP & R & 718.9 & 65.7 &   23.4 $\pm$ \phantom{0}5.1 &   12.1 $\pm$ \phantom{0}5.1  &   26.3 $\pm$   \phantom{0}5.1 &   13.7 $\pm$   \phantom{0}5.5\\
2017AUG & 2017-08-07 11:01:51 & 2081 & 1280 & 650LP & R & 718.9 & 65.7 &   30.9 $\pm$ \phantom{0}5.1 &   21.5 $\pm$ \phantom{0}5.2 &   37.6 $\pm$   \phantom{0}5.1 &   17.4 $\pm$   \phantom{0}3.9\\
2018JUL & 2018-07-23 10:38:00 & 955 & 640 & 650LP & R & 722.3 & 64.8 &   27.5 $\pm$ \phantom{0}5.8 &   33.6 $\pm$ \phantom{0}5.8 &   43.4 $\pm$   \phantom{0}5.8 &   25.4 $\pm$   \phantom{0}3.8\\
\hline
2005APR & 2005-04-27 & $-$ & $-$ & BRB & APD & 753.8 & 93.0 &   11.1 $\pm$   \phantom{0}3.0 &   20.6 $\pm$ \phantom{0}3.0 &   23.4 $\pm$   \phantom{0}3.0 &   30.8 $\pm$   \phantom{0}3.6\\
\hline
\hline
\end{tabular}
\begin{flushleft}
{$^a$} APD means `avalanche photodiode'; B, R indicate blue- and red-sensitive H10720-210 and H10720-20 PMTs, respectively.
\end{flushleft}
\end{table*}

\begin{table*}
\caption{Observations of interstellar control stars}
\label{tab:controls}
\tabcolsep 3.5 pt
\begin{tabular}{rlllrrrrrrrr}
\hline
\hline
\multicolumn{1}{c}{Control}      & SpT   &   Run     & UT                  & Dwell & Exp. & $\lambda_{eff}$ & Eff. & \multicolumn{1}{c}{$q$} & \multicolumn{1}{c}{$u$} & \multicolumn{1}{c}{$p$} & \multicolumn{1}{c}{$\theta$}\\
\multicolumn{1}{c}{HD} &        &            &           & \multicolumn{1}{c}{(s)} & \multicolumn{1}{c}{(s)} & \multicolumn{1}{c}{(nm)} & \multicolumn{1}{c}{(\%)} & \multicolumn{1}{c}{(ppm)} & \multicolumn{1}{c}{(ppm)} & \multicolumn{1}{c}{(ppm)} & \multicolumn{1}{c}{($^\circ$)}\\
\hline
165777 & A5V & 2017AUG & 2017-08-12 14:16:03 & 1409 & 640 & 468.3 & 87.9 &   14.9 $\pm$   \phantom{0}6.3 &   41.4 $\pm$   \phantom{0}6.6 &   44.0 $\pm$   \phantom{0}6.5 &   35.1 $\pm$   \phantom{0}4.1 \\
171802 & F5III & 2018AUG & 2018-08-19 10:23:03 & 1533 & 1120 & 469.0 & 73.1 &   11.3 $\pm$   \phantom{0}8.4 &   27.6 $\pm$   \phantom{0}7.6 &   29.8 $\pm$   \phantom{0}8.0 &   33.9 $\pm$   \phantom{0}7.8 \\
173880 & A5III & 2017AUG & 2017-08-18 11:30:53 & 1916 & 640 & 466.6 & 87.4 &    6.0 $\pm$   \phantom{0}9.4 &   12.5 $\pm$   \phantom{0}9.3 &   13.9 $\pm$   \phantom{0}9.4 &   32.2 $\pm$  24.0 \\
175638 & A5V & 2018AUG & 2018-08-18 10:29:58 & 991 & 640 & 464.4 & 70.5 &   $-$2.2 $\pm$   \phantom{0}7.4 &   50.6 $\pm$   \phantom{0}6.9 &   50.6 $\pm$   \phantom{0}7.2 &   46.2 $\pm$   \phantom{0}4.2 \\
181391 & G8/K0IV & 2018AUG & 2018-08-19 15:22:44 & 978 & 640 & 475.9 & 77.1 &  $-$27.7 $\pm$  10.3 &    3.1 $\pm$  10.0 &   27.9 $\pm$  10.2 &   86.8 $\pm$  11.5 \\
182640 & F1IV-V(n) & 2018MAR & 2018-04-04 18:44:03 & 1075 & 640 & 467.4 & 82.6 &  $-$22.5 $\pm$   \phantom{0}4.2 &    6.0 $\pm$   \phantom{0}4.3 &   23.3 $\pm$   \phantom{0}4.2 &   82.5 $\pm$   \phantom{0}5.2 \\
\hline
\hline
\end{tabular}
\begin{flushleft}
Notes: * All control star observations were made with the SDSS g$^\prime$ filter and the B PMT as the detector. The same aperture as used for the $\alpha$~Oph observations in the same run was used. * Spectral types are from SIMBAD, as is all position and distance information presented later.\\
\end{flushleft}
\end{table*}

A previously reported high-precision polarimetric observation of $\alpha$~Oph made with PlanetPol \citep{bailey10} is used in this work. That observation was made in April 2005 at the 4.2-m William Herschel Telescope (WHT) located at the Roque de los Muchachos Observatory on the island of La Palma in the Canary Islands. PlanetPol utilised photo-elastic modulators (PEMs) to provide rapid modulation in a single broad red photometric band (BRB) and is described by \citet{hough06}.

Between September 2016 and April 2019 we also made 29 new high-precision polarimetric observations of $\alpha$~Oph in seven photometric bands with a variety of instrument and telescope configurations. The bulk of the observations were made using HIPPI (HIgh Precision Polarimetric Instrument, \citealp{bailey15}) and its successor HIPPI-2 \citep{bailey19b} at the 3.9-m Anglo-Australian Telescope (AAT) at Siding Spring Observatory. Two observations were made at the Gemini North telescope on Mauna Kea with HIPPI-2. A single observation was made at UNSW Observatory in Sydney with a 35-cm Celestron C14 telescope using Mini-HIPPI \citep{bailey17}. The standard operating procedures for each instrument were followed, with the reduction procedures described for HIPPI-2 by \citet{bailey19b} used to process all of the data. 

The common characteristic of HIPPI-class polarimeters is their use of Ferro-electric Liquid Crystal (FLC) modulators to provide primary modulation at a frequency of 500~Hz. This enables the limitations induced by atmospheric seeing to be overcome and results in parts-per-million precision. The ultimate precision limit is a set-up and wavelength-dependent positioning error associated with how accurately the target is centred in the aperture. These errors have been determined for all the instrumental configurations used here and are given in \citet{bailey19b}. Total errors are determined as the quadratic sum of the positioning error and the internal measurement error, which is a function of exposure time.

A second stage of modulation, to remove instrumental effects, is accomplished by rotating either the telescope Cassegrain rotator (HIPPI) or the instrument rotator (HIPPI-2, Mini-HIPPI) in turn through four position angles, $PA=0, 45, 90, 135^{\circ}$. HIPPI had a third stage of modulation whereby the back-end including the detectors was rotated 90$^{\circ}$ approximately every 20 to 40~s. A single sky (S) measurement was made adjacent to each target (T) measurement at each of the four position angles, $PA=0, 45, 90, 135^{\circ}$, in the pattern TSSTTSST. On a number of occasions observations made in different filters were made back-to-back.

All the observations used either blue-sensitive Hamamatsu H10720-210 or red-sensitive Hamamatsu H10720-20 photo-multiplier tubes (PMTs) as the detectors. 

For the HIPPI-2 observations our standard filter set, described in \citet{bailey19b}, was used; briefly, this includes 425- and 500-nm short-pass filters (425SP, 500SP), SDSS $g^{\prime}$, Johnson V, SDSS $r^{\prime}$ and a 650~nm long-pass (650LP) filter. With HIPPI, Omega Optics versions of the SDSS $g^{\prime}$ and $r^{\prime}$ filters were used instead of the Astrodon versions used with HIPPI-2. The blue-sensitive PMTs were paired with most of the filters, with the red-sensitive PMTs used for the 650LP observations, and for some of the $r^{\prime}$ observations. 

Other set-up features impact the bandpasses. Different versions of the FLC modulator were used, for one of which the performance has drifted over time, with the performance at different eras calibrated independently. HIPPI-2 was designed for an f/16 focus, but when used at the AAT f/8 focus a 2$\times$ negative achromatic (Barlow) lens was used which attenuates more light at shorter wavelengths; the effect is appreciable in the 425SP filter. In combination with the photometric filters the passbands had effective wavelengths ($\lambda_{\rm eff}$) between 399~nm and 723~nm (the PlanetPol observation has a slightly longer $\lambda_{\rm eff}$, 754~nm). The instrument/telescope configurations details for each run are summarised in Table~\ref{tab:runs}. 

A small polarization due to the telescope mirrors, TP, results in shifts to the zero-point offset of our observations. These are corrected for by reference to the straight mean of several observations of low-polarization standard stars, 
details of which are given either in \citet{bailey17, bailey19b} or in the caption of Table~\ref{tab:runs}. Similarly, the position angle ($\theta$) -- measured eastward from celestial north -- is calibrated by reference to literature measurements of high-polarization standards, also given in either \citet{bailey17, bailey19b} or in the caption of Table~\ref{tab:runs}.
The observations of $\alpha$~Oph are given in Table~\ref{tab:observations}; here the positioning error is included in the reported uncertainties. The data are presented in order of effective wavelength, with horizontal lines used to group similar band passes. The demarcation in the 425SP data separates out observations that made use of the Barlow lens from those that didn't. Coincidentally, the observations with the Barlow lens coupled with modulator performance that was less efficient at short wavelengths. As the modulator-efficiency curve becomes increasingly steep as it decreases at bluer wavelengths this implies a lesser weighting to the shortest wavelengths than accounted for by the effective wavelength -- which does not take account of modulation efficiency. 

\begin{figure}
\includegraphics[width=8.25cm]{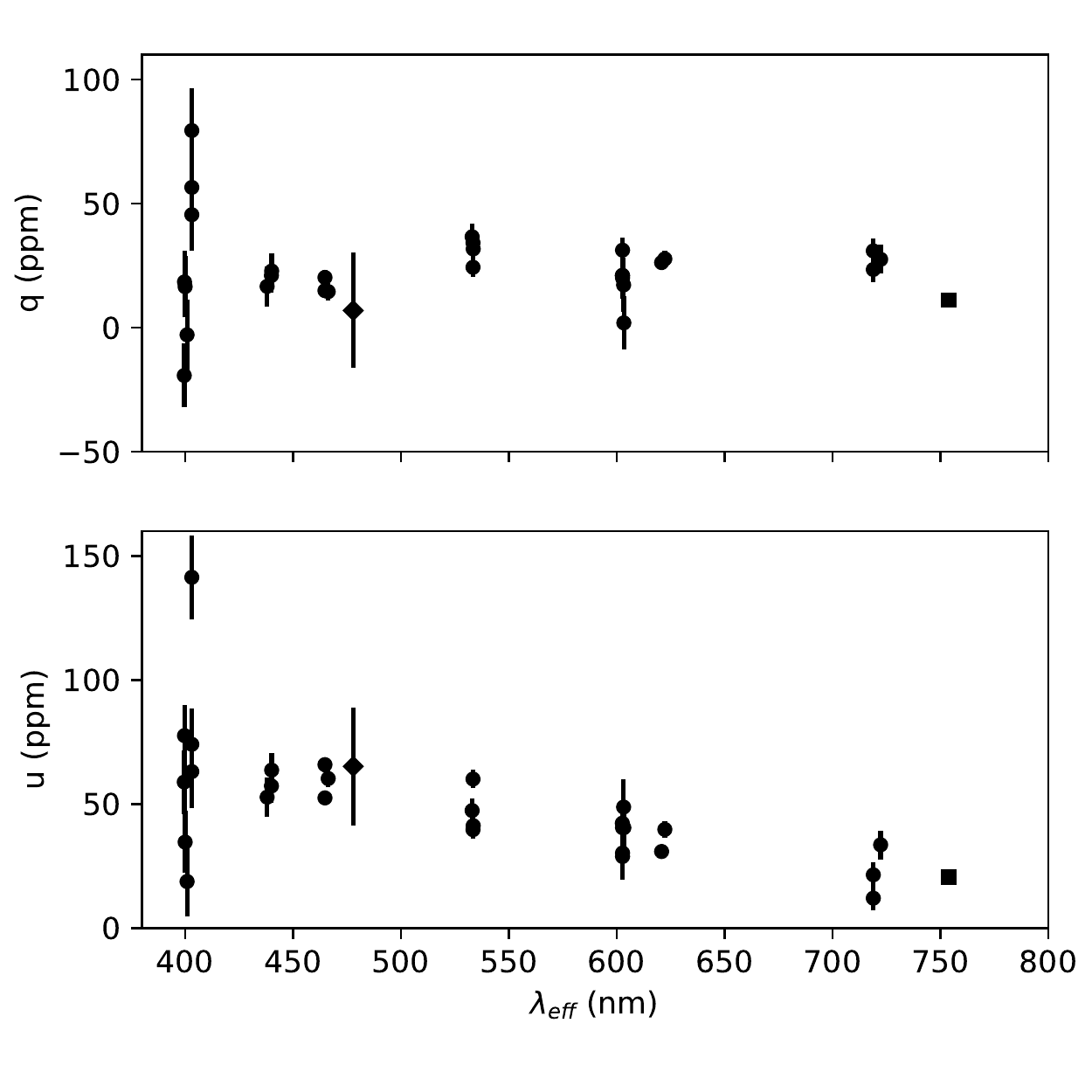}
\caption{Observations made of $\alpha$~Oph with HIPPI and HIPPI-2 (circles), Mini-HIPPI (diamond, at 477.9~nm) and PlanetPol (square, at 753.8~nm).}
 \label{fig:observations}
\end{figure}

The observational data, in terms of $q=Q/I$ and $u=U/I$, are also shown graphically in Fig.~\ref{fig:observations}, where a trend with wavelength is clear in both Stokes parameters, but particularly in $u$ where the polarization is higher, and reminiscent of those shown in Fig.~\ref{fig_pol_95}. The decline in polarization with wavelength from the central maximum is steeper than can be accounted for by interstellar polarization, which is characterised by a Serkowski curve \citep{serkowski75}. However, some component of the polarization measured must be due to the interstellar medium.

A common way of gauging the magnitude and orientation of interstellar polarization is to observe intrinsically unpolarized control stars that are near to the object on the sky and at a similar distance \citep{clarke10}. We have previously found stars with spectral types ranging from A to early K to be the least intrinsically polarized \citep{cotton16a, cotton16b}. Such stars are a good probe of the nearby interstellar medium so long as debris-disk hosts \citep{cotton17b} and active stars \citep{cotton19a} are avoided. 

A number of suitable control stars are to be found in the \textit{Interstellar List} in the appendix of \citet{cotton17b}. We have supplemented the Interstellar List with new observations reported here in Table~\ref{tab:controls} for the first time. The interstellar control stars were observed in the same fashion as $\alpha$~Oph using the SDSS g$^\prime$ filter, and the reported precision includes the positioning error.

\section{Modelling}
\label{sec_model}

The modelling approach described here is essentially the same as used for our study of Regulus \citep{cotton17}, but is presented here with a little more detail.

\subsection{Rotating-star geometry}
\label{sec_geom}

\begin{figure*}
    \centering
    \includegraphics[width=12cm]{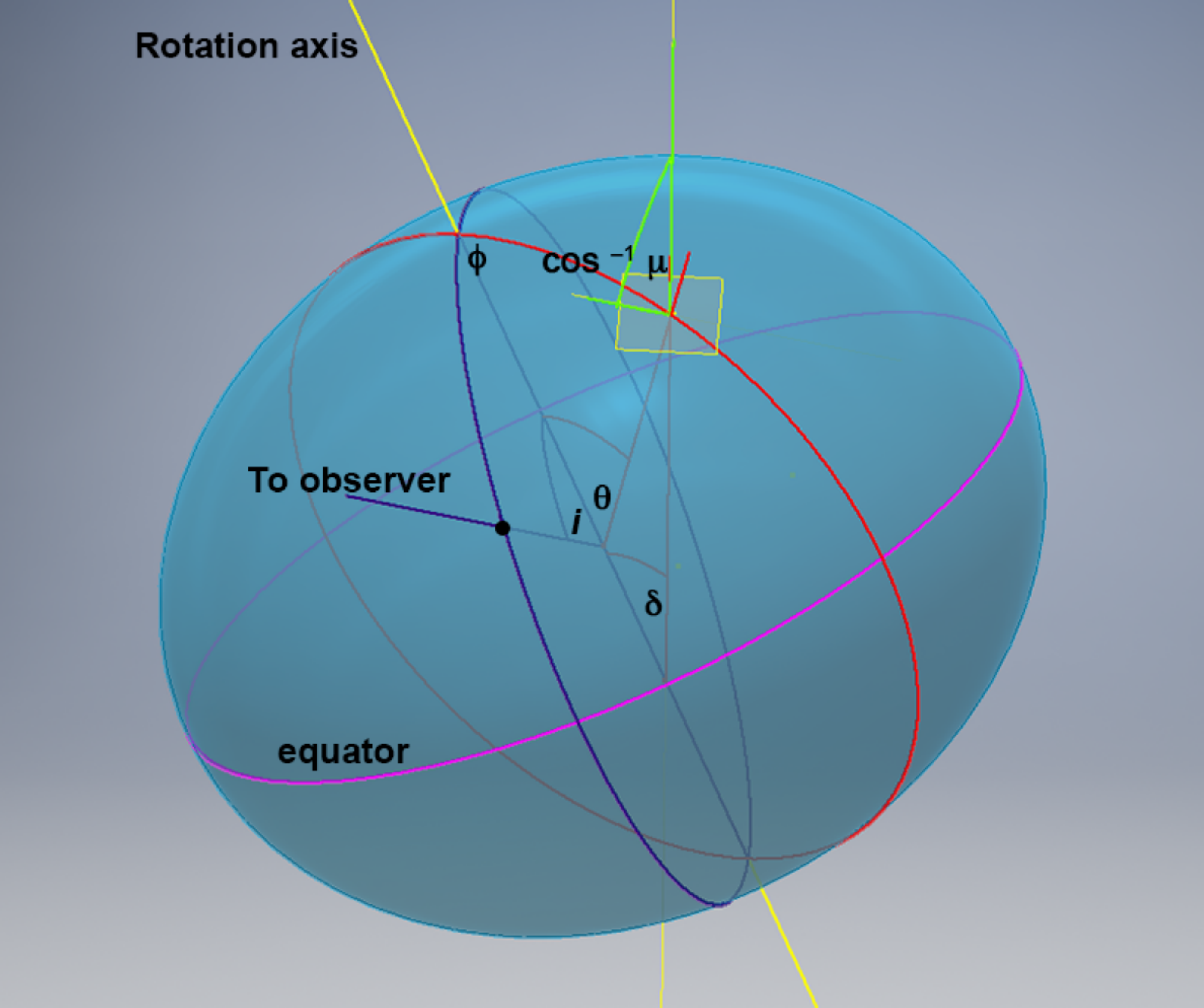}
    \caption{Coordinate system for rotating star models as used in section~\ref{sec_geom}. The blue meridian is towards the observer and defines the zero of longitude ($\phi$). The square is a tangent plane to the surface at longitude $\phi$ and colatitude $\theta$. The normal to this tangent plane (which is in the direction of the local effective gravity) makes an angle $\delta$ with the rotation axis and an angle $\cos^{-1} (\mu)$ with the direction to the observer. Lines in each different plane are indicated by different colours. The angle $\xi$ defined in equation~(\ref{eqn:xi}) is the angle between the `green' plane and the `blue' plane.}
    \label{fig:geom}
\end{figure*}

We assume the Roche model for a rotating star, in which the mass is concentrated at the centre. We define $\omega$ = $\Omega/\Omega_{\rm c}$ as the angular velocity of the star (assumed to be uniformly rotating) divided by the critical angular velocity, given by:
\begin{equation}
    \Omega_{\rm c} = \sqrt{\frac{8GM}{27R_{\rm p}}}
\end{equation}
where $M$ is the mass of the star and $R_{\rm p}$ its polar radius. For this standard Roche model, the shape of the rotationally distorted star is completely specified by the $\omega$ parameter, and is given by
\begin{equation}
x(\theta,\omega) = \frac{3}{\omega \sin{\theta}} \cos{[1/3(\pi+\cos^{-1}({\omega \sin{\theta}}))]}
\end{equation}
 \citep{harrington68}, where $x$ is the radius of the star at colatitude $\theta$ in units of the polar radius.  The local normal to the star's surface is at an angle $\delta$ to the rotation axis of the star, given by:
\begin{equation}
\tan{\delta} = \left(1 - \frac{8}{27}x^3\omega^2\right) \tan{\theta}.
\end{equation}
If the star is viewed with the rotation axis at an inclination angle $i$ to the line of sight then a point on the star's surface at longitude $\phi$ and colatitude $\theta$ will be seen at a viewing angle $\mu$ (cosine of the observer's local zenith distance), where
\begin{equation}
    \mu = \sin{i}\sin{\delta}\cos{\phi} + \cos{i}\cos{\delta}.
\end{equation}
See Fig.~\ref{fig:geom} for further explanation of the coordinate system.

For polarimetric modelling we also need the rotation angle $\xi$ \citep[][and see Fig.~ \ref{fig:geom}]{harrington68} given by:
\begin{equation}
    \xi = \tan^{-1}\left(\frac{\sin{\delta}\sin{\phi}}{\sin{i}\cos{\delta}-\cos{i}\sin{\delta}\cos{\phi}}\right)
    \label{eqn:xi}
\end{equation}
which is required to rotate polarization vectors from a local atmospheric radiative-transfer solution to the co\"ordinate system of the observer.

\subsection{Gravity darkening}

A rotating star has gravity and temperature varying over its surface. The variation of gravity over the surface is defined by the Roche model \citep[e.g.][]{collins66}. The variation of temperature depends on a gravity-darkening model. We use the gravity-darkening model of \citet{espinosa11} which is specifically designed for rotating stars. It results in slightly less gravity darkening than predicted by the traditional \citet{vonzeipel24} law, which has the form $T_{\rm eff} \propto g^{\beta}$ with $\beta$ = 0.25. 

The effective-temperature profile of the star as a function of colatitude is calculated using equations 24 and 31 of \citet{espinosa11} and is dependent on the rotation rate ($\omega$). We note that these authors use a different definition of $\omega$ than the one adopted here. They define $\omega_{\rm elr} = \Omega/\Omega_k$ where $\Omega_k$ is the Keplerian angular velocity at the star's actual equatorial radius, rather than the critical angular velocity $\Omega_{\rm c}$. This gravity-darkening law is close to the von~Zeipel law for slowly-rotating stars, but for rapid rotators it results in less variation of temperature for a given change in gravity, in agreement with interferometric results that indicate values of $\beta$ less than 0.25 for a number of rapidly-rotating stars \citep{monnier07,zhao09,che11}.

\subsection{Stellar-atmosphere models}

The emergent-flux distribution of a rotating-star model can be specified by three parameters:\footnote{Assuming locally plane-parallel geometry, and for given abundances and microturbulent velocities. Other parameter sets can provide equivalent information; e.g., the equatorial temperature, or the global effective temperature, can be substituted for the polar value. An additional parameter is required to set the overall flux scaling (most transparently the polar radius, although the equatorial rotation speed or the stellar mass provide equivalent information in the Roche formulation).   The \textit{observed} flux additionally depends on the axial inclination $i$ (plus the distance and interstellar extinction).} the rotation rate, $\omega$, together with $T_{\rm p}$ and $g_{\rm p}$, the local, polar effective temperature and gravity. Given these parameters, we calculate the emission from rotating-star models by dividing the model surface into a large number of tiles, and interpolating specific intensities  (spectral radiances) for each tile as functions of $g, T, \lambda\text{, and } \mu$.

The intensities come from pre-computed grids of \textsc{atlas9} solar-composition stellar-atmosphere models;  for the polarization calculations discussed in Section~\ref{sec_radtran}, these are tailored to the local effective temperature and gravity for colatitudes from 0$\degr$ to 90$\degr$ at 2$\degr$ intervals, based on \citet{castelli03} models, while the parameter-space explorations presented in Section~\ref{sec:grid} use equivalent results from \citet{howarth11}.

\subsection{Polarized radiative transfer}
\label{sec_radtran}

To calculate the specific intensity and polarization for each of the 46 tailored models we use a modified version of the \textsc{synspec} spectral synthesis code \citep{hubeny85,hubeny12} in which we have replaced the standard radiative transfer with a fully polarized radiative-transfer calculation, using the \textsc{vlidort} code of \citet{spurr06}. \textsc{vlidort} (Vector Linearized Discrete Ordinate Radiative Transfer) is a comprehensive implementation of the discrete-ordinate method of radiative transfer. It has been widely used in Earth-atmosphere applications but is equally suitable for astronomy and there are a number of examples of its application to planetary and stellar atmospheres \citep{kopparla16,cotton17,bailey18,bott18,bailey19a}.

To incorporate \textsc{vlidort} into \textsc{synspec} we use \textsc{synspec}'s determination of the absorption, emission and scattering properties of each atmospheric layer at each wavelength to derive the vertical optical depth, single scattering albedo and equivalent black body emission which are the inputs required by \textsc{vlidort} for each atmospheric layer. We include scattering from electrons and Rayleigh scattering from H, He and H$_2$. All these scattering processes are assumed to polarize light according to a Rayleigh scattering matrix, and all line and continuum absorption processes are treated as pure absorption with no effect on polarization.

The outputs from \textsc{synspec/vlidort} are intensity and polarization values for each of the 46 $\theta$ values, modelled as functions of viewing angle ($\mu$) and wavelength. The wavelength is on a non-uniform scale chosen by \textsc{synspec} to fully sample the spectral line structure; we rebin the data to a uniform 0.01-nm wavelength spacing for subsequent analysis.

These methods have been extensively tested. \textsc{vlidort} has itself been tested against a number of benchmark problems in polarized radiative transfer \citep{spurr06}. \citet{bailey18} used \textsc{vlidort} to reproduce classic results on the polarization phase curves of Venus from \citet{hansen74}, showing excellent agreement both with observations and with the original calculations that used a quite different radiative-transfer method (the doubling method). In \citet{cotton17} we verified polarization and intensity calculations for hot stars using \textsc{synspec/vlidort} against earlier results from \citet{harrington15}, again showing good agreement.

\subsection{Integration over the star}
\label{subsec_int}

For a given inclination, we overlay the `observed' view of the model star with a rectangular pattern of pixels spaced at 0.01 times the polar radius. For each pixel that overlaps the star we determine the longitude ($\phi$) and colatitude ($\theta$) on the surface and then use the relationships from section~\ref{sec_geom} to determine the corresponding viewing angle $\mu$ and rotation angle $\xi$. Given this information we then interpolate in our grid of intensity and polarization results obtained in section~\ref{sec_radtran} to obtain the intensity and polarization for this pixel. Linear interpolation is used first in $\theta$ and then in $\mu$. The polarization vectors (described by Stokes $Q$ and $U$ values) are then rotated through 2$\xi$ to put all the values into the reference frame of the pixel grid.

\begin{figure*}
    \centering
    \includegraphics[angle=90,width=14cm]{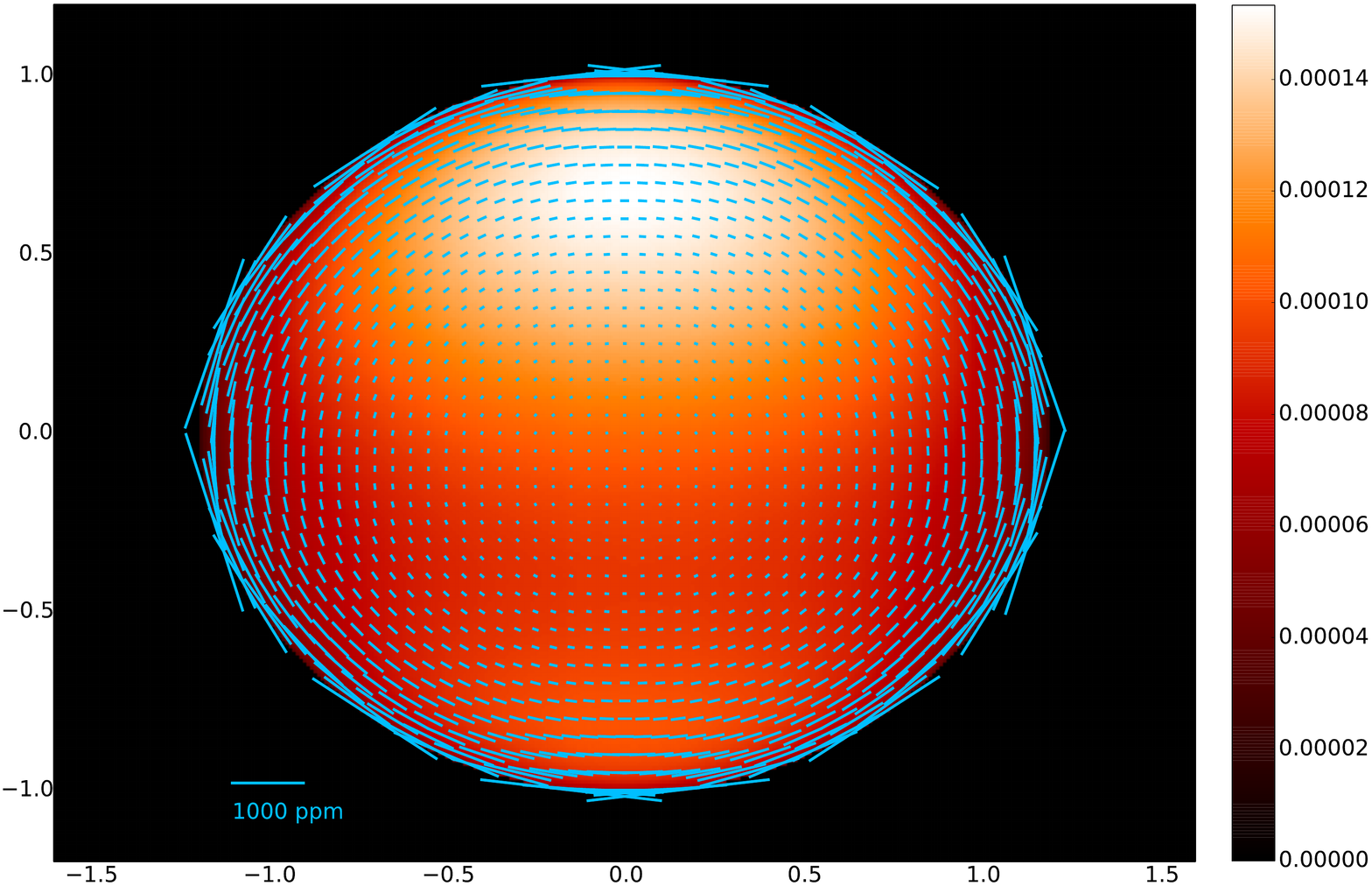}
    \caption{Polarization vectors overlaid over intensity distribution for a model of Rasalhague with $\omega= 0.88$ and inclination = 75$\degr$ at a wavelength of 400~nm. The intensity scale is specific intensity (or spectral radiance) $I_\nu$ in units of erg~cm$^{-2}$~s$^{-1}$~Hz$^{-1}$~sr$^{-1}$.}
    \label{fig:pol_image}
\end{figure*}

The resulting pixel values can be plotted as an image of the intensity distribution with overlaid polarization vectors, as in Fig.~\ref{fig:pol_image}. The intensity and polarization can also be summed over all pixels to give the integrated polarization for the star as a function of wavelength, which can be directly compared with observations.

In \citet{cotton17} we tested the full modelling procedure described above by comparison with past calculations of the wavelength dependent polarization of rotating B stars by \citet{sonneborn82} and found good agreement.

\subsection{Additional Constraints}
\label{sec:addc}

The predicted polarization depends primarily on four model parameters: the rotation rate $\omega$, the (e.g., polar) gravity and temperature ($g_{\rm p}$, $T_{\rm p}$), and the axial inclination ($i$). As found for Regulus by \citep{cotton17}, it is not possible to determine these four parameters uniquely by modelling  polarization data alone. We therefore use additional observational information that provides further relationships between these parameters. 

A first constraint comes from the observed spectral-energy distribution, which is principally sensitive to the effective temperature (the global $T_{\rm eff}$, or the polar $T_{\rm p}$).  Knowledge of the distance then establishes the overall linear scale of the system (e.g., polar radius). The projected equatorial rotation speed, $v_{\rm e}\sin{i}$, obtained from spectroscopy, then determines $\omega$, for given $i$ and $g_{\rm p}$. These constraints therefore provide relationships between the four polarization-sensitive parameters such that, for any given rotation rate $\omega$ and inclination $i$, we can determine the corresponding $g_{\rm p}$ and $T_{\rm p}$ values, as well as a number of related parameters (such as the mass).

Note that we do not use the interferometric data as an additional constraint. While interferometric measurements provide information on the rotation for $\alpha$~Oph \citep{zhao09} and some other stars, we want to develop a method of measuring rotation that is independent of interferometry and can potentially be applied to stars that are too faint or too distant for such methods.

In the case of $\alpha$~Oph, a number of published measurements of $v_{\rm e}\sin{i}$ exist \citep{slettebak75,gray80,carpenter84,abt95,erspamer03}. These measurements are in reasonable agreement, averaging $212 \pm 8$~km~s$^{-1}$ (range: 198 to 220~km~s$^{-1}$). We additionally carried out our own analysis of data from the Elodie archive \citep{prugniel04}. Using spectral synthesis of the Mg\,II~$\lambda$4481 line, for a full gravity-darkened Roche model matching our final parameters,  we find $v_{\rm e}\sin{i} = 230 \pm 5$~km~s$^{-1}$ (from both direct and fourier-transform modelling). This is in good agreement with the value of $228 \pm 4$~km~s$^{-1}$ reported by \citet{royer02}, which is a recalibration of the data from \citet{abt95}. 

\begin{figure}
    \centering
    \includegraphics[width=\columnwidth]{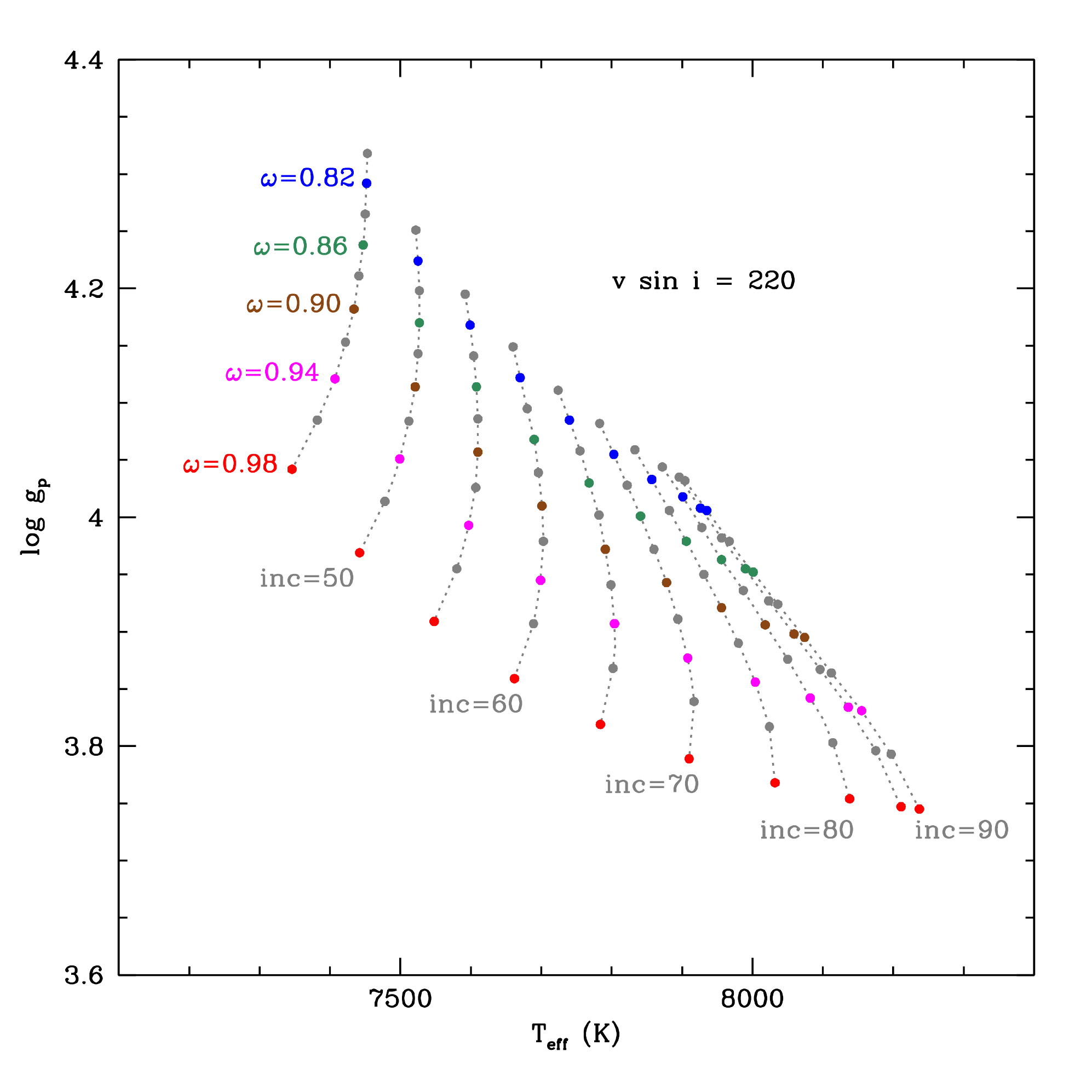}
    \caption{Example of the model grids described in section~\ref{sec:grid}. The relationship between the grid values of $\omega$ and inclination and the star's (global) effective temperature and polar gravity is shown here for $v_{\rm e}\sin{i}$ = 220~km~s$^{-1}$. (For given $\omega$, $i$, the global effective temperature, determined by the overall luminosity, corresponds to a specific polar temperature $T_{\rm p}$.)}
    \label{fig:grid}
\end{figure}

\begin{figure*}
\includegraphics[width=17.7cm]{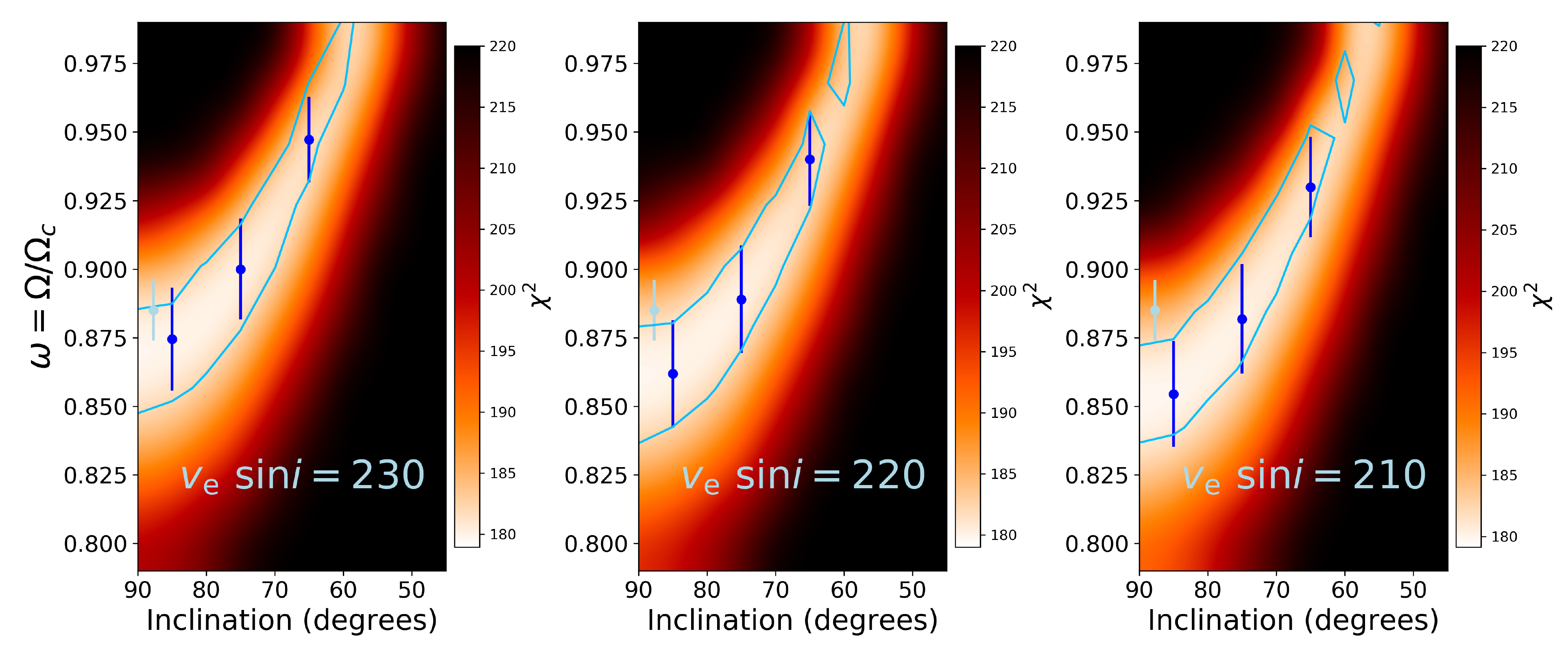}
\caption{$\chi^2$ fits to polarization of $\alpha$~Oph. As described in section~\ref{sec:grid} all positions on these grids correspond to model stars that match the observed spectrum. The $\chi^2$ values show additionally which of these models are consistent with the observed polarization. The blue error bars show the 1 $\sigma$ uncertainties in $\omega$ at selected inclinations as determined using the bootstrap procedure described in section~\ref{sec:uncertainties}. The corresponding 1 $\sigma$ contours are also shown. The light blue point with error bars is the interferometric determination from \citet{zhao09}.}
\label{fig:fits}
\end{figure*}

The observed spectral-energy distribution is based on photometry from \citet{johnson66} and archival low-resolution IUE spectra, from which we derive an integrated flux over 1210--3010~\AA\ of $3.39 \times 10^{-7}$~erg~cm$^{-2}$~s$^{-1}$. The parallax is $67.13 \pm 1.06$~milliarcsec \citep{vanLeeuwen07} and the spectral type is A5IV \citep{gray01}. The extinction is assumed to be negligible, based on the very low polarization reported by \citet{bailey10} and in Section~\ref{sec_bfm}.

\subsection{Model Grids}
\label{sec:grid}

We make use of the constraints described above to develop grids of parameters to use in the polarization modelling. For a spherical star the observed flux is primarily dependent on the effective temperature and the radius. However, for a rapidly-rotating star there is additionally a dependence on the rotation ($\omega$) and inclination ($i$). The procedure is therefore to start with assumed values of $\omega$, $i$ and $v_{\rm e}\sin{i}$ and then determine the $T_{\rm eff}$, $R_{\rm p}$ values for which the modelled fluxes reproduce the observed spectrum. 

We use a simple interval-halving method to determine the polar radii that match the observed M$_V$ for a range of $T_{\rm eff}$ values, giving a locus in the $T_{\rm eff}/R_{\rm p}$ plane. The procedure is then repeated to match the integrated UV flux, giving a second locus. For temperatures relevant to B-/A-type stars these two loci are quite distinct, and their intersection provides the well-determined $T_{\rm eff}/R_{\rm p}$ pair consistent with the adopted $\omega$, $i$ and $v_{\rm e}\sin{i}$ values.

For $\alpha$~Oph we constructed grids at $v_{\rm e}\sin{i}$ values of 210, 220, and 230~km~s$^{-1}$ (section~\ref{sec:addc}), covering $\omega$ values from 0.8 to 0.98 in steps of 0.02 and inclinations from 45$\degr$ to 90$\degr$ in 5$\degr$ steps. The grid for $v_{\rm e}\sin{i} = 220$~km~s$^{-1}$ is illustrated in Fig.~\ref{fig:grid}. Each point in the grid corresponds to a model star that reproduces the adopted $v_{\rm e}\sin{i}$ and observed spectrum of the star. By comparing the predicted polarization to that observed we can further constrain the stellar parameters. We use the methods described in sections \ref{sec_geom} to \ref{subsec_int} to calculate the predicted polarization as a function of wavelength for each point in the model grid as shown.

\begin{figure}
    \centering
    \includegraphics[width=\columnwidth]{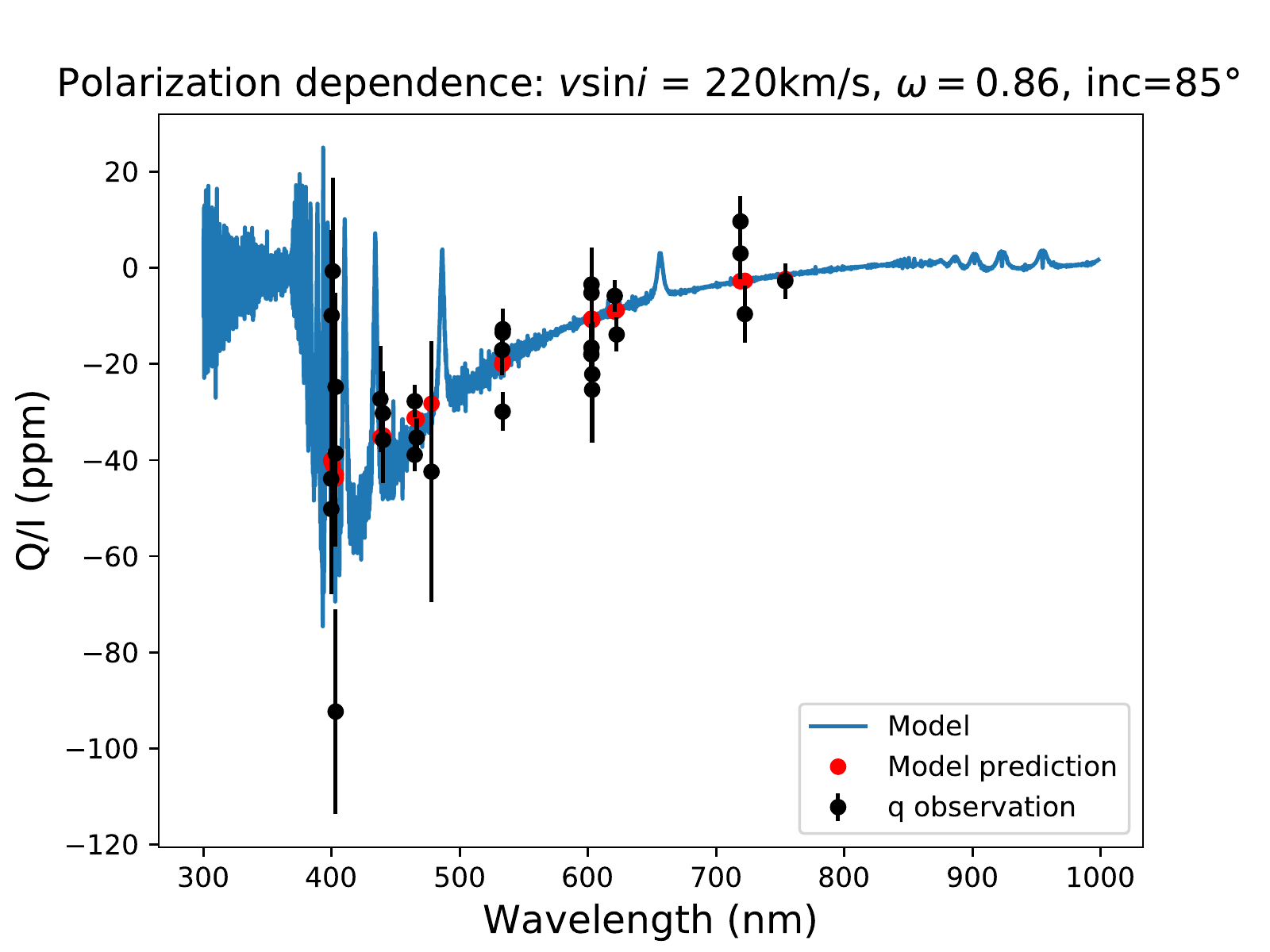}
    \includegraphics[width=\columnwidth]{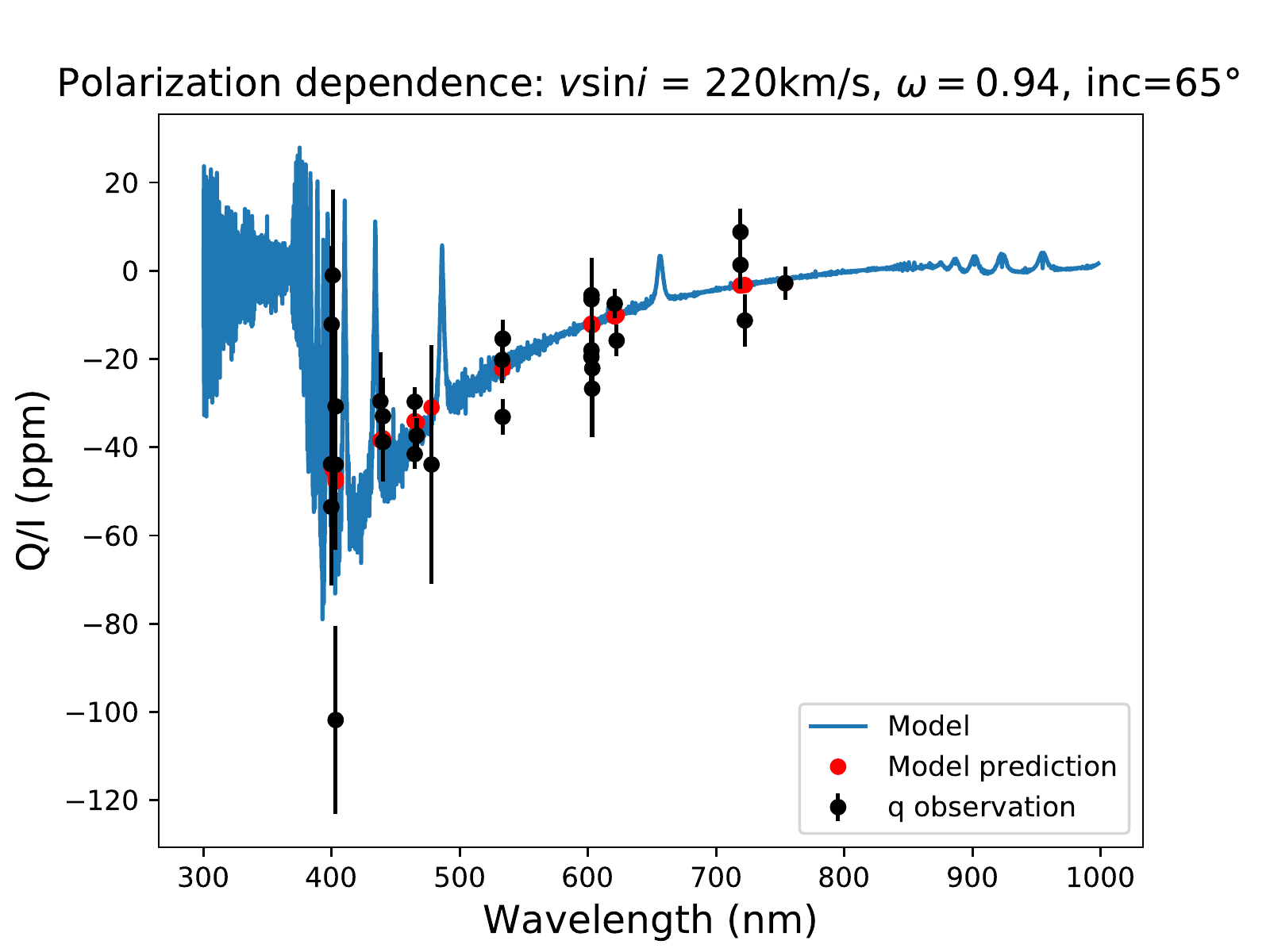}
    \caption{Modelled polarization wavelength dependence (blue line) for two of the best-fitting grid models compared with the observations corrected for interstellar polarization. Red points are the model prediction integrated over the filter bandpasses for each observation.}
    \label{fig:mod_wave}
\end{figure}

\section{Results and Discussion}
\label{sec_discuss}

\subsection{Comparison of observed and predicted polarization}

From the observations presented in section~\ref{sec_obs}, we have measurements of the $q$ and $u$ Stokes parameters for a range of wavelengths (Fig.~\ref{fig:observations}). From the modelling described in section~\ref{sec_model} we have a prediction of the polarization as a function of wavelength for each of the model-grid points described in section~\ref{sec:grid}.

We aim to find the model or models in the grid that best-fit the observations. There are a number of complications to this process. First, the filters used in HIPPI and HIPPI-2 are relatively broad. Thus to determine the model polarization for a filter we integrate the predicted polarization over each filter bandpass, rather than simply using the modelled value corresponding to the effective wavelength of the filter. This is facilitated by the detailed bandpass model for the HIPPI instruments described by \citet{bailey19b}.

Secondly, the polarization due to the rotating star will depend on the position angle of the star's rotation axis, which, in general, is unknown. The measured $q$ and $u$ are relative to celestial north. Because the star is symmetric with respect to its rotation axis, the orientation of its intrinsic polarization must be either parallel or perpendicular to its rotation axis. Hence we need to apply a rotation to the observed $q$ and $u$ Stokes parameters which has the effect of putting all the polarization into the $q$ parameter, while the $u$ parameter should be zero within measurement uncertainties.

The third complication is interstellar polarization, described in the next section.

\subsection{Interstellar polarization}
\label{interstellar}

In addition to any intrinsic polarization due to rotation, a star will also show some interstellar polarization. Interstellar polarization has a distinctive wavelength dependence described by the Serkowski law \citep{serkowski75} which, as updated by \citet{wilking82} and \citet{whittet92}, has the form:
\begin{equation}
    p(\lambda) = p_{\rm max} \exp{[-K \ln^2 (\lambda_{\rm max}/\lambda)]}
    \label{eqn:serk}
\end{equation}
where
\begin{equation}
    K = 0.01\pm0.05 + (1.66\pm0.09)\lambda_{\rm max}.
\end{equation}

While we made some observations of interstellar control stars to get an estimate of the interstellar polarization for $\alpha$~Oph as described in section~\ref{sec_obs}, the scatter in these observations is such that they can only be used as a rough guide to the expected level. Therefore we determine the interstellar polarization parameters as part of the fitting procedure. 

The method relies on the fact that the wavelength dependence of polarization due to rotation (e.g. Fig.~\ref{fig_pol_95}) is very different from the wavelength dependence of interstellar polarization as defined by equation~(\ref{eqn:serk}).   For each model in the grid we determine the differences between the observed polarization and the predicted rotating-star polarization and fit a Serkowski curve as defined in equation~(\ref{eqn:serk}) to these differences. The fit is carried out using the \textsc{curve\_fit} routine of the \textsc{python} package \textsc{scipy} \citep{scipy}. There are four fit parameters, $p_{\rm max}$, $\lambda_{\rm max}$, $\theta$ (the position angle of the interstellar polarization) and PA$_{\rm rot}$ (the rotation-axis position angle of the star).

\begin{table*}
    \centering
    \caption{Best-fit $\omega$ and 1$\sigma$ limits on $\omega$ as a function of inclination.} 
        \begin{tabular}{|l|ll|ll|ll|}
        \hline
          & \multicolumn{2}{c|}{$v_{\rm e}\sin{i} = 210 $}  & \multicolumn{2}{c|}{$v_{\rm e}\sin{i} = 220 $} & \multicolumn{2}{c|}{$v_{\rm e}\sin{i} = 230 $} \\
        inclination & $\omega$ & 1 $\sigma$ range & $\omega$ & 1 $\sigma$ range & $\omega$ & 1 $\sigma$ range  \\
        \hline
         60$\degr$ & 0.958 & 0.942 -- 0.975 & 0.965 &  0.949 -- 0.980 & 0.973 & 0.956 -- 0.980 \\
         65$\degr$ & 0.929 & 0.909 -- 0.949 & 0.940 &  0.911 -- 0.956 & 0.943 & 0.926 -- 0.962 \\
         70$\degr$ & 0.904 & 0.884 -- 0.920 & 0.911 &  0.891 -- 0.926 & 0.916 & 0.898 -- 0.935 \\
         75$\degr$ & 0.882 & 0.862 -- 0.900 & 0.889 &  0.869 -- 0.909 & 0.900 & 0.878 -- 0.902 \\
         80$\degr$ & 0.865 & 0.845 -- 0.887 & 0.876 &  0.854 -- 0.896 & 0.885 & 0.861 -- 0.902 \\
         85$\degr$ & 0.855 & 0.836 -- 0.875 & 0.862 &  0.844 -- 0.884 & 0.875 & 0.853 -- 0.893 \\
         90$\degr$ & 0.851 & 0.833 -- 0.872 & 0.860 &  0.840 -- 0.880 & 0.867 & 0.849 -- 0.893 \\
         \hline
    \end{tabular}

    \label{tab:tab_omeg}
\end{table*}

\begin{figure}
    \centering
    \includegraphics[clip, trim= 0cm 0.6cm 0cm 0cm, width=\columnwidth]{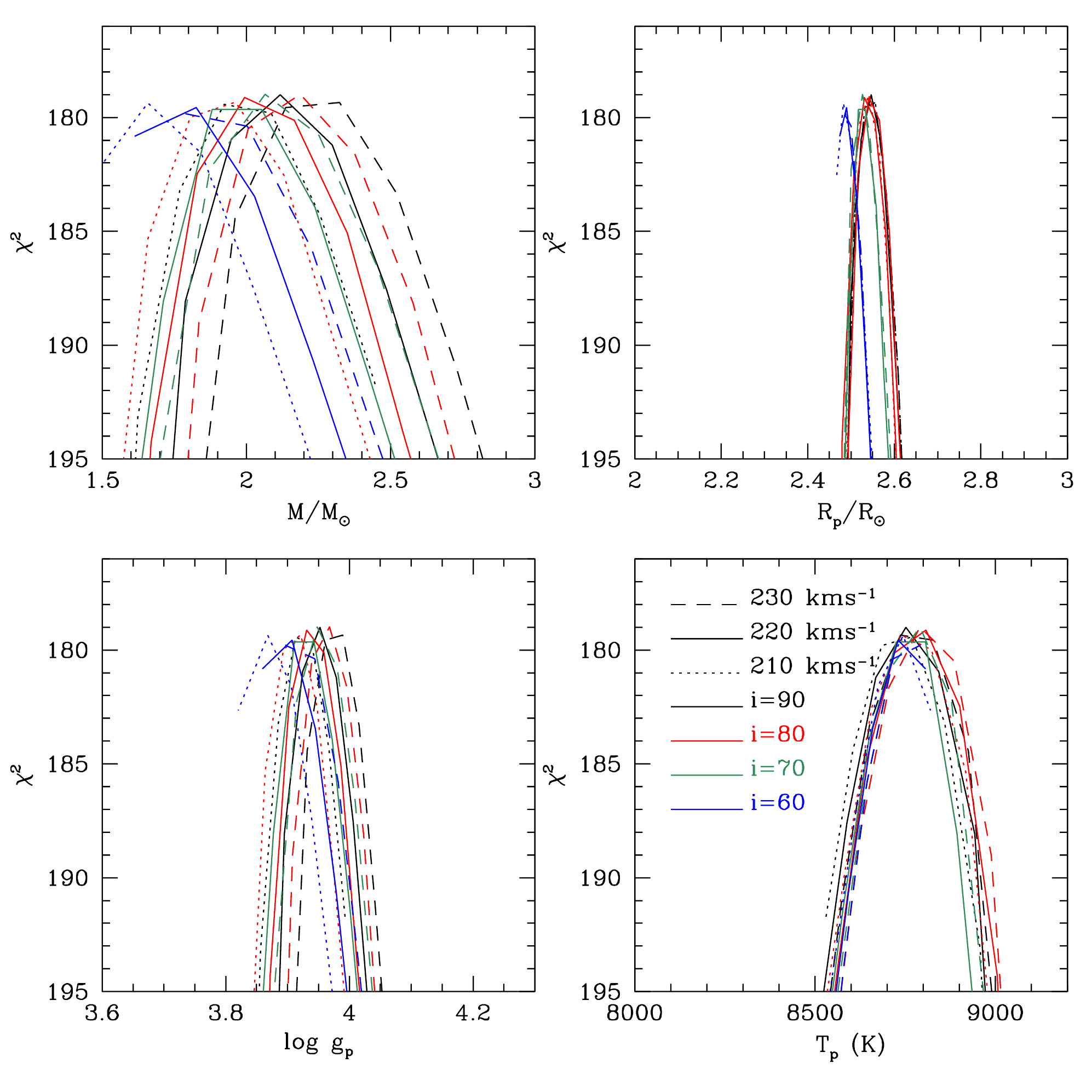}
    \caption{Stellar parameters for $\alpha$~Oph plotted against $\chi^2$ values as shown in Fig.~\ref{fig:fits}}
    \label{fig:star_chi}
\end{figure}

\begin{figure}
    \centering
    \includegraphics[clip, trim = 0cm 0.6cm 0cm 0.4cm, width=\columnwidth]{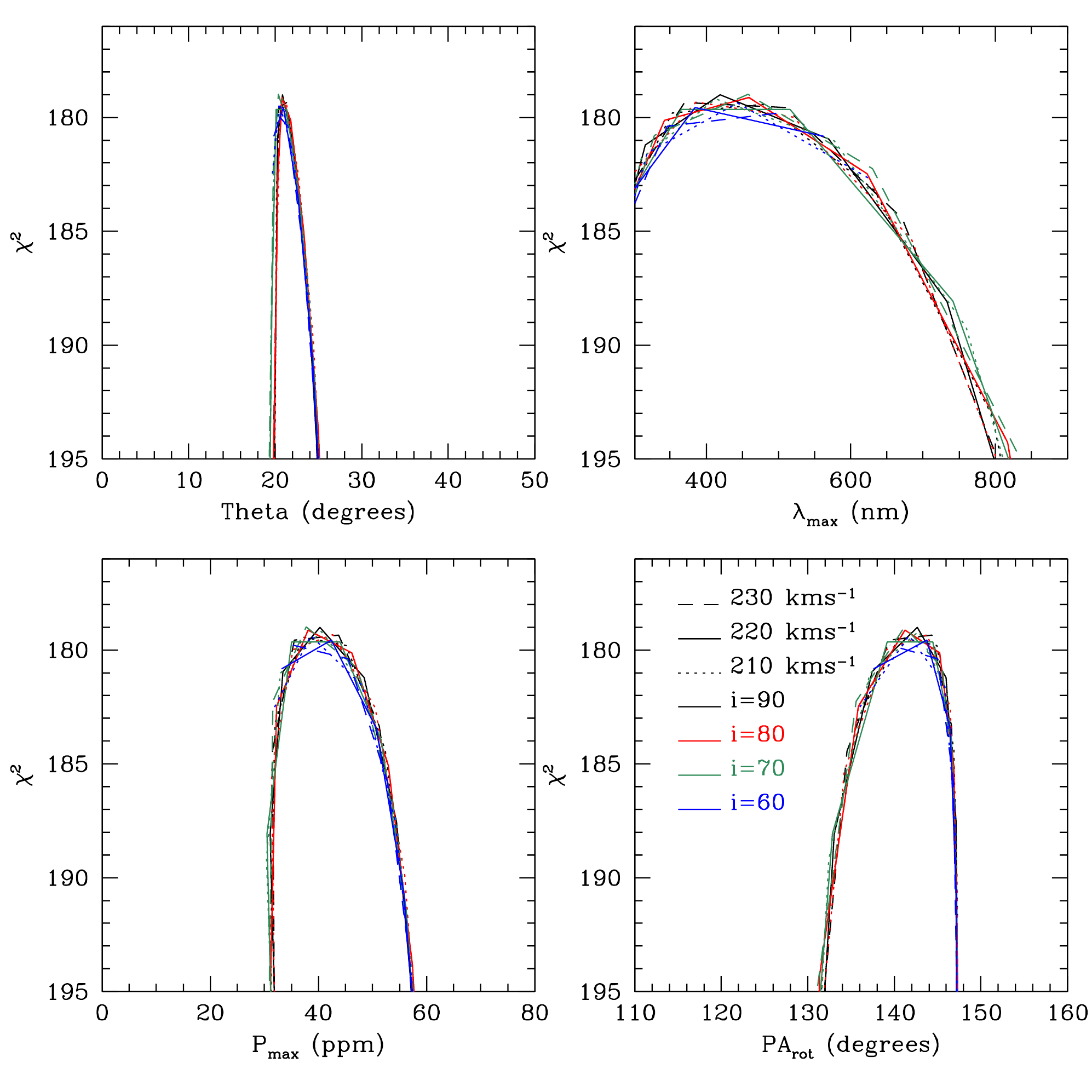}
    \caption{Interstellar parameters and stellar rotation axis for $\alpha$~Oph plotted against $\chi^2$ values as shown in Fig.~\ref{fig:fits}}
    \label{fig:int_chi}
\end{figure}

\subsection{Best-fit models}
\label{sec_bfm}

Having established best-fit values for these four fit parameters at each model grid point we can then determine the $\chi^2$ value describing the deviation of the observations from the model. The results of this procedure for our three model grids are shown in Fig.~\ref{fig:fits}. When we applied the analogous procedure to Regulus \citep[Fig.~4]{cotton17} we were able to constrain  $\omega$ and $i$ to a relatively small part of the diagram. It can be seen that the results are less constraining in the case of $\alpha$~Oph. There is a diagonal region across the diagram where good fits to the polarimetry are obtained covering a range of $\omega$ from $\sim$0.83 to $\sim$0.98 and inclinations from $\sim$60$\degr$ to $\sim$90$\degr$. 

There are two reasons the parameters are less constrained. First, in the case of Regulus, the interstellar polarization was very small, and we assume $\lambda_{\rm max}$ and determined $p_{\rm max}$ and $\theta$ from the control-star measurements; fitting these values necessarily introduces additional uncertainty. The second reason relates to the different nature of the polarization curves for hotter B stars like Regulus and for A~stars like $\alpha$~Oph, as can be seen in Fig.~\ref{fig_pol_95}. At higher temperatures the polarization changes sign from positive to negative in the middle of the wavelength range. The wavelength at which this crossover occurs is temperature and inclination dependent and therefore helps to reduce the degeneracy in parameters and separate the effects of $\omega$ and inclination. In the cooler stars the polarization is always negative and the effects of $\omega$, inclination and gravity on the polarization curve are all similar, making these parameters harder to constrain. 

\begin{table}

    \centering
    \caption{Stellar and interstellar parameters for $\alpha$~Oph} 
        \begin{tabular}{lll}
        \hline
        Parameter & Value (all $i$) &  Value ($i$=90) \\
        \hline
        \multicolumn{2}{l}{Stellar Parameters} \\
        \hline
        $\omega$  & 0.905 $\pm$ 0.075 &   0.863  $\pm$ 0.033\\  
        $i$ (deg)  &    75  $\pm$ 15   &   90    \\
        PA$_{\rm rot}$ (deg)  &   142 $\pm$ 4  &  142 $\pm$ 4 \\
        $\log{g_{\rm p}}$  &   3.93  $\pm$  0.08 &  3.95 $\pm$ 0.06 \\
        $T_{\rm p}$ (K) & 8725 $\pm$ 175 & 8695 $\pm$ 145 \\
        $T_{\rm eff}$ (K) & 7855 $\pm$ 205 & 8010 $\pm$ 50 \\
        $R_{\rm p}$ (\rsun)  &  2.52 $\pm$ 0.06  &  2.53 $\pm$ 0.04 \\
        $\log{L}$ (\lsun) & 1.455 $\pm$ 0.025 & 1.471 $\pm$ 0.009 \\
        $M$ (\msun)   &  2.0 $\pm$ 0.4  &  2.1 $\pm$ 0.3 \\
        \hline
        \multicolumn{3}{l}{Interstellar Polarization Parameters}  \\
        \hline
        P$_{\rm max}$ (ppm) &  40.2 $\pm$ 7.1 & 40.3 $\pm$ 6.9  \\
        $\lambda_{\rm max}$ (nm) &  440 $\pm$ 110 &  419 $\pm$ 110  \\
        $\theta$  (deg)   &   21.0  $\pm$ 1.1  &  20.9  $\pm$ 1.1\\
        \hline
        \end{tabular}
    \label{tab:parameters}
\end{table}

Fig.~\ref{fig:mod_wave} shows the modelled polarization wavelength dependence compared with the observed polarization, corrected for the interstellar contribution, and rotated to put all the intrinsic polarization into the $Q$ Stokes parameter. Two examples of best-fitting combinations of $\omega$ and $i$ are shown, but the models look almost the same for any of the best-fitting combinations traced out in Fig.~\ref{fig:fits}.

\subsection{Uncertainties}
\label{sec:uncertainties}

The uncertainties on our best-fit parameters were determined by a bootstrap procedure. There are 30 observations of $\alpha$~Oph used in our analysis. In the bootstrap procedure we constructed random sets of observations by drawing 30 times from our observation set with replacement, such that an individual observation could be selected several times, or not at all. We repeated the fitting process described above in 1000 such trials, creating 1000 different versions of the results shown in Fig.~\ref{fig:fits}. We determined the best-fitting $\omega$ values at a range of inclinations (using spline interpolation to allow for the grid resolution) and determined the 1-$\sigma$ errors from the statistics of the trials. We could then determine the $\Delta \chi^2$ corresponding to 1$\sigma$, which we found to be 1.9 (rather than the value of one that would be expected if the model was perfect, and the errors on the observations were correctly estimated). This $\Delta \chi^2$ can then be used to determine error bounds on $\omega$, $i$, or any of the other stellar parameters that are related to them through the grid constraints (section~\ref{sec:grid}) or the fitting procedure.

\subsection{Stellar parameters}

Table~\ref{tab:tab_omeg} lists the best-fitting $\omega$ and 1$\sigma$ limits for each modelled inclination from 60$^\circ$ to 90$^\circ$, as determined from the data plotted in Fig.~\ref{fig:fits}. Models with inclination less than 60$^\circ$ fall outside the 1-$\sigma$ error bounds as described above. There is a strong correlation between $\omega$ and inclination, as can be seen from Fig.~\ref{fig:fits}. Using our grid constraints we are also able to determine values for many of the other stellar parameters. Results for these are listed in Table~\ref{tab:parameters}. We list two sets of parameters. The second column (`all $i$') represents the range of values if we allow any inclination, while the third column is the range of values if we consider only the $i = 90^\circ$ results.

\citet{zhao09} have given an interferometric analysis of the rotation of $\alpha$~Oph. They obtain $\omega = 0.885 \pm 0.011$ and $i = 87.70^\circ \pm 0.43^\circ$ for a model with \citet{vonzeipel24} gravity darkening (i.e. $\beta = 0.25$). These values are reasonably consistent with our results, particularly for the $v_{\rm e}\sin{i}$ = 230~km~s$^{-1}$ models (\citeauthor{zhao09} derive $v_{\rm e}\sin{i}$ = 237~km~s$^{-1}$ from their analysis.) If the interferometric determination of the inclination is correct then the results we obtain for high inclination (third column of Table~$\ref{tab:parameters}$) will be the most relevant ones.

However, the rotation-axis position angle we obtain, $142^\circ \pm 4^\circ$, is significantly different from that obtained by \citet{zhao09};  they give $-53.88^\circ \pm 1.23^\circ$, equivalent to $126.12^\circ$ (since linear polarimetry cannot distinguish $180^\circ$ differences in angle). This is a $16^\circ$ or $4\sigma$ difference from our result. $\alpha$~Oph is a rapidly-rotating star in a binary system, so precession of its rotation axis should occur. The observations used by \citet{zhao09} were made in 2006 and 2007, while the bulk of our polarization observations were made over 2017--2019. (The PlanetPol observation, made in 2005, should not effect our rotation-axis determination as the rotational polarization is near zero at this red wavelength.) Thus the possibility that the position angle is changing due to precession should be considered and could be tested by future observation. If precession is occurring then the inclination could be changing as well as the position angle, but if the discrepancy is not due to time variations then there must be another factor that is impacting on the position-angle measurement. For example, there may be another source of polarized light in the system that we have not included in our analysis. If that is the case then other parameters we have determined here might also be affected. However, it seems unlikely that there could be another source that would exactly mimic the distinctive polarization wavelength dependence due to rotation, and therefor our general conclusions should remain valid.

Our flux-model grids, described in section~\ref{sec:grid}, allow us to set constraints on a number of other stellar parameters by looking at the distribution of the corresponding $\chi^2$ values, as shown in Fig.~\ref{fig:star_chi}. Results, listed in Table~\ref{tab:parameters}, include a polar gravity  $\log{g_{\rm p}} = 3.93 \pm 0.08$, polar radius $R_{\rm p} = 2.52 \pm 0.06~\rsun$, and polar temperature  $T_{\rm p} = 8725 \pm 175$~K. These values are a little different from those obtained by \citet{zhao09}, but we note that their model is based on a different gravity-darkening law (von~Zeipel) compared with that of \citet{espinosa11} used in our analysis.

The stellar mass from our analysis is $M = 2.0 \pm 0.4~\msun$. A dynamical mass of $2.40^{+0.23}_{-0.37}~\msun$ has been obtained from analysis of the binary orbit \citep{hinkley11} while \citet{zhao09} give a mass of $2.10 \pm 0.02~\msun$ from comparison with evolutionary models (of non-rotating stars). Our result is consistent with these values.

\begin{figure*}
    \centering
    \includegraphics[clip, trim = {1.4cm 1.0cm 2.5cm 2.5cm}, width=\columnwidth]{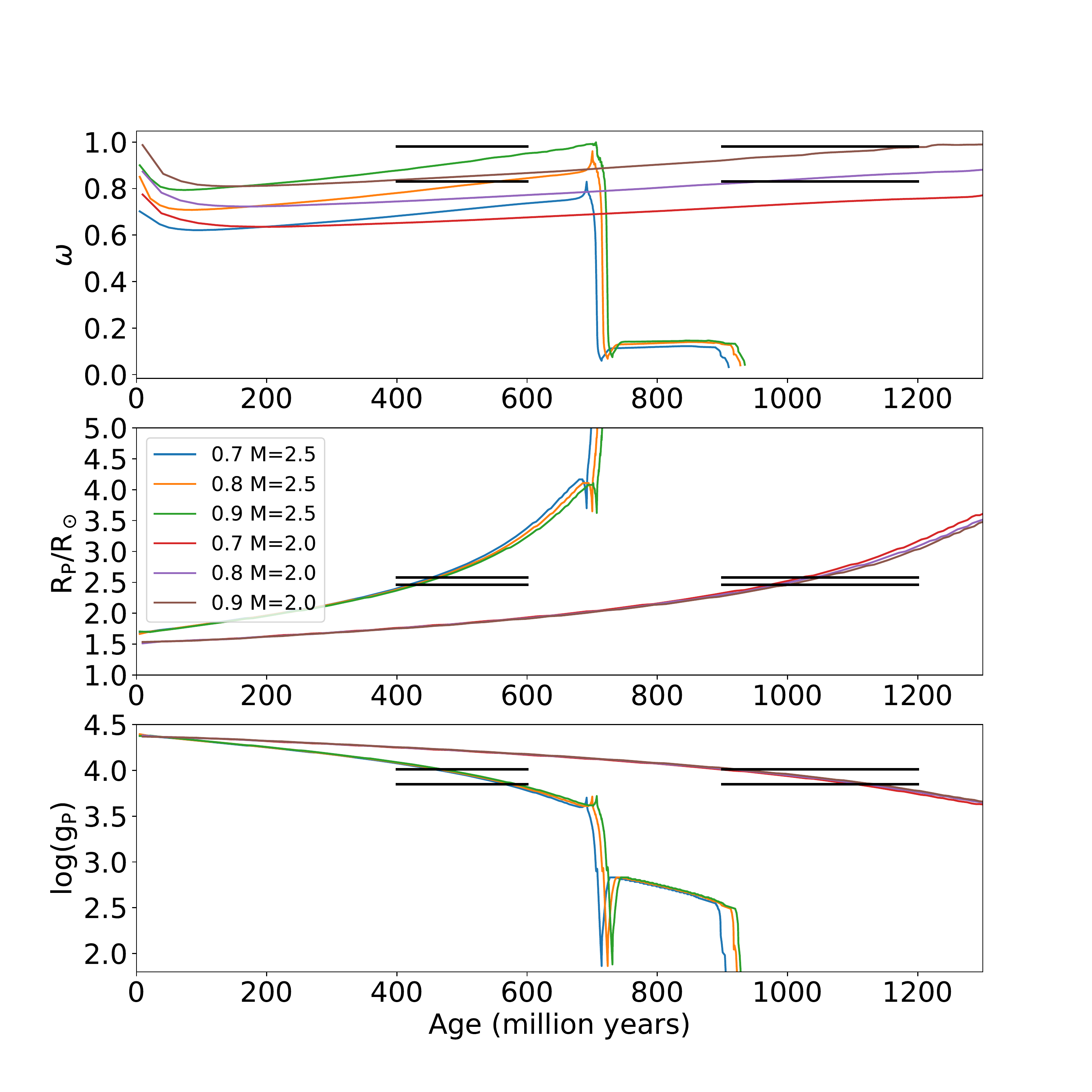}
    \includegraphics[clip, trim = {0.4cm 0.7cm 2.5cm 2.5cm},width=\columnwidth]{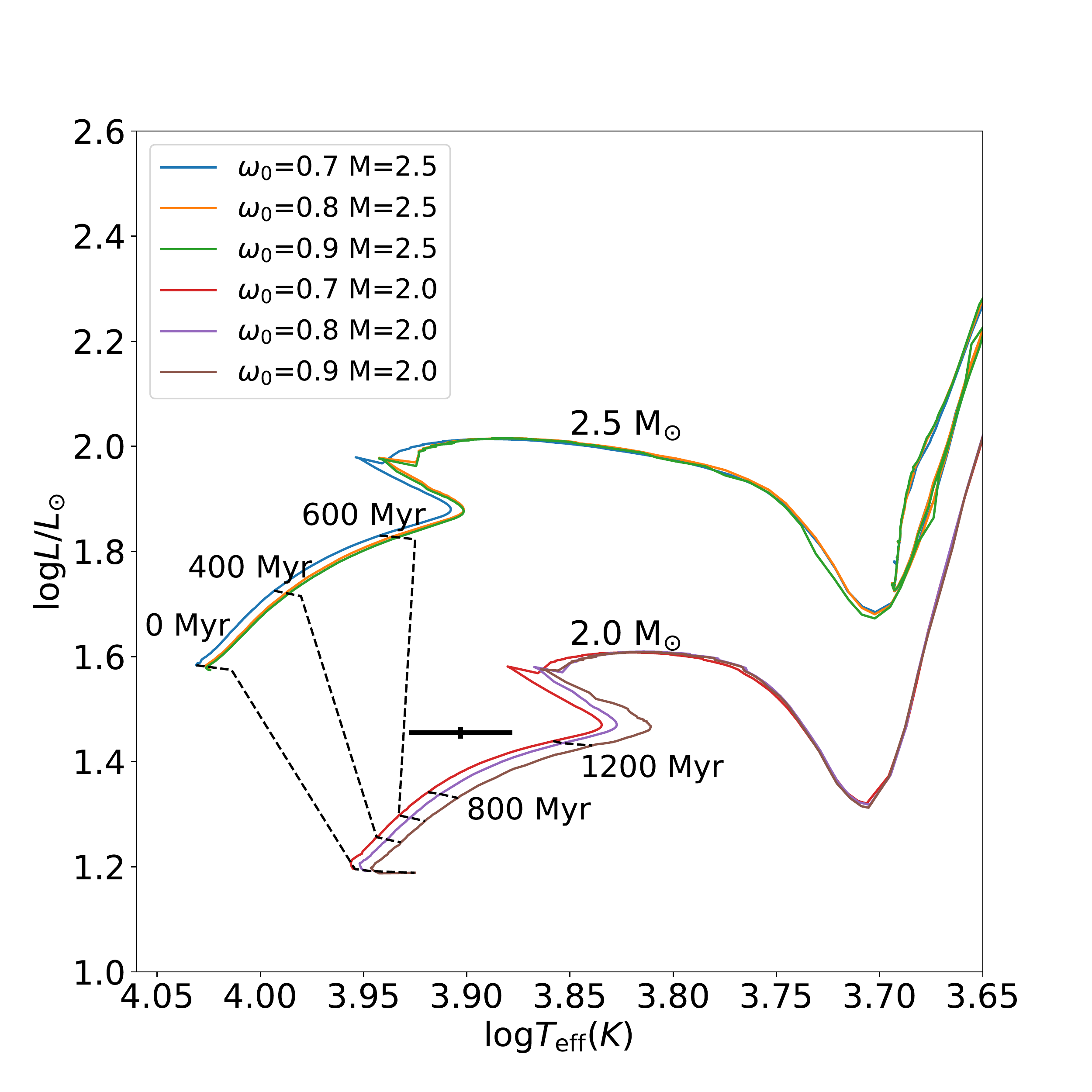}
    \caption{Evolutionary models for rotating stars with $M=2.5\msun$ and 2.0$\msun$, $Z=0.014$, and initial rotation rates of $\omega$ = 0.7, 0.8, and 0.9, from \citet{georgy13}. On the left rotation rate, polar radius and gravity are plotted against time;  horizontal black lines show the range of values indicated by modelling of our observations. On the right the models are plotted on an HR diagram with our measured luminosity and $T_{\rm eff}$ shown.}
    \label{fig:evol}
\end{figure*}

\subsection{Evolutionary state}

The most recent spectral classification of $\alpha$~Oph is A5IVnn \citep{gray01}, but in the the past it has been assigned luminosity classifications from III \citep{levato72} to V \citep{levato78}. The spectroscopically determined luminosity classification for rapidly-rotating stars can be misleading as an indicator of evolutionary status, because of the reduced equatorial gravity due to rotation. In Fig.~\ref{fig:evol} we show a comparison of our measured rotation, polar radius and polar gravity with the evolutionary-model predictions for rotating stars of mass 2.5~$\msun$ and 2.0~$\msun$ \citep{georgy13}. The gravity, in particular, constrains $\alpha$~Oph to be in the later half of its main-sequence evolution. For this 2.0 $\msun$ model the indicated age is $\sim$900--1200 million years or $\sim$400-600 million years for a 2.5~$\msun$ model. The initial rotation rate of the star must have been greater than $\omega \simeq 0.8$. On the HR diagram (right panel of Fig.~\ref{fig:evol}) our measured luminosity and $T_{\rm eff}$ place $\alpha$~Oph nearer the 2.0~$\msun$ track and again in the later part of the main sequence evolution.

\subsection{Interstellar parameters}

\begin{figure*}
\includegraphics[clip, trim={2.7cm 0cm 2.5cm 1cm}, width=18cm]{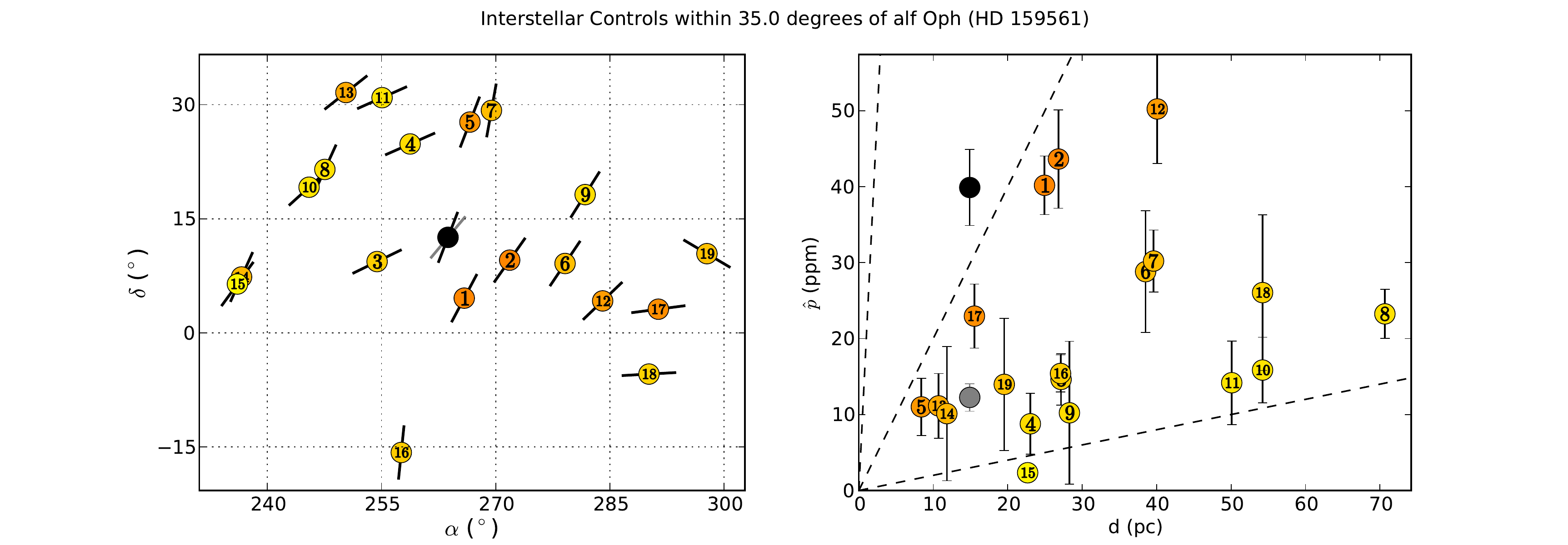}
\caption{A map (left) and p vs d plot (right) of interstellar control stars within 35$^\circ$ of $\alpha$~Oph. Interstellar PA ($\theta$) is indicated on the map by the black pseudo-vectors; and defined as the angle North through East, i.e. increasing in a clockwise direction with vertical being 0$^\circ$. The controls are colour coded in terms of $\hat{p}/d$ and numbered in order of their angular separation from $\alpha$~Oph; they are: 1: HD~161096, 2: HD~165777, 3: HD~153210, 4: HD~156164, 5: HD~161797, 6: HD~171802, 7: HD~163993, 8: HD~148856, 9: HD~173880, 10: HD~147547, 11: HD~153808, 12: HD~175638, 13: HD~150860, 14: HD~141004, 15: HD~140573, 16: HD~155125, 17: HD~182640, 18: HD~181391, 19: HD~187691. In the p vs d plot dashed lines corresponding to $\hat{p}/d$ values of 0.2, 2.0 and 20.0 ppm/pc are given as guides. The grey data-point is derived from the interstellar model in \citep{cotton17b} and the black  data-point represents our best-fit interstellar values for $\alpha$~Oph (converted to 450~nm to compare with the g$^\prime$ observations using the Serkowski Law assuming $\lambda_{\rm max}$=~440~nm).}
 \label{fig:is_map}
\end{figure*}

In addition to the stellar parameters, a number of interstellar-polarization parameters are also presented in Table~\ref{tab:parameters}. As can be seen from Fig.~\ref{fig:int_chi} the determinations of these parameters are largely model independent, and so they are almost the same in the ``all $i$'' and ``$i = 90$'' columns. Since little rotationally-induced polarization is modelled for the longest-wavelength bands, these measurements play a disproportionate role in their determination. The interstellar polarization is an obscuring element for the primary science in this paper, but these parameters have scientific value in their own right. Studies of interstellar polarization reveal the structure of the interstellar magnetic field, as well as the properties and history of dust in the ISM \citep{clarke10, frisch15, jones15}. 

Of particular interest is $\lambda_{\rm max}$ since, given $\alpha$~Oph's distance from the Sun of only $\sim$15~pc, our investigation samples the most local ISM yet studied in this manner. $\lambda_{\rm max}$ is inversely proportional to the particle size of dust in the interstellar medium \citep{draine95}. A value of 550~nm is typical for our Galaxy \citep{serkowski75} and in the rim of the Local Hot Bubble \citep{cotton19b} -- a region of space largely devoid of gas and dust, carved out by ancient supernovae, that extends roughly 75 to 150~pc beyond the Sun \citep{liu16}. 

Only two prior measurements of $\lambda_{\rm max}$ have been made within the Local Hot Bubble. \citet{cotton19b} found $\sim350 \pm50$~nm by making multi-band polarization measurements of four stars between 47 and 92~pc distant, each thought to be intrinsically unpolarized, and fitting the Serkowski--Wilking Law with \citet{whittet92}'s value for $K$ (see section~\ref{interstellar}). Similarly, \citet{marshall16} used two-band measurements of five stars between 19 and 27~pc to infer that $\lambda_{\rm max}$ was approximately 470~nm (but could be between 35 and 600~nm). The value determined here for $\alpha$~Oph of 440~$\pm$~110~nm is consistent with these prior results, and with grains shocked by the evolution of the Loop~I Superbubble \citep{cotton19b}.

In Fig.~\ref{fig:is_map} the values of $\theta$ and $P_{\rm max}$ are compared with measurements made of nearby stars that have been selected for probably being intrinsically unpolarized. As outlined in section~\ref{sec_obs}, these stars either come from the \textit{Interstellar List} in \citet{cotton17b} or are new control observations. In the left-hand panel of Fig.~\ref{fig:is_map} it can be seen that the polarization position angles of stars near to $\alpha$~Oph fairly consistently point to around 45$^\circ$. This is reflected in the grey pseudo-vector, which is a separation-weighted average of the controls determined according to \citet{cotton17b} 
\begin{equation} 
Wt = (1 - s_a/35),
\end{equation} where $s_a$ is the separation to $\alpha$~Oph in degrees. The difference between this average polarization position angle and the one determined by our best-fit-models method (section~\ref{sec_bfm}) is 19.3~$\pm$~1.1$^\circ$. Such a difference is what one might expect from a pair of stars separated by $\sim$7.5$^\circ$ on the sky \cite[][Fig.~5]{cotton17b}. We can't expect an agreement much better than this, so this gives some confidence in the results of the fitting procedure.

For stars with galactic latitudes $-90^\circ <b<+30^\circ$, the magnitude of interstellar polarization is given by \citet{cotton17b} as 
\begin{equation}
p_i = (1.644 \pm 0.298)(d - 14.5) + (11.6 \pm 1.7),
\end{equation} 
which for $\alpha$~Oph ($d=$~14.9~pc, $b = +22.6^\circ$) gives $p_i=12.3~\pm~1.8$~ppm, as represented by the grey circle in the right-hand panel of Fig.~\ref{fig:is_map}. By contrast, the black circle represents the value of $P_{\rm max}$ found using the best-fit method (and adjusted for wavelength); it is roughly three times higher. The interstellar-polarization magnitude we find for $\alpha$~Oph is also higher than any of the other surrounding control stars in terms of $p/d$. However, two of the stars with the largest values for $p/d$ within this nearby volume of space are also the two closest to $\alpha$~Oph (labelled 1 and 2 in Fig.~\ref{fig:is_map}). The difference is approximately 1.5-$\sigma$/0.8~ppm/pc, but this is much smaller than the scatter in the other control stars. So, while the value of $P_{\rm max}$ is higher than expected, it is not unreasonably so.

In \citet[][Supplementary Materials]{cotton17} we used the models of \citet{cotton17b} rather than fitting the interstellar parameters. Regulus is in a region of the sky with a smoother dust distribution and lower dust content; $p_i$ was estimated to be only 6.3~ppm, so this was a reasonable approach to take. However, with a larger interstellar polarization for $\alpha$~Oph this would not have been sufficient. If our best-fit models were to fit only $PA_{rot}$, $\chi^2_r$ would be three to four times worse, and the other determined stellar parameters somewhat different. This underlines the difficulty in using interstellar controls for precision polarimetry in nearby space.

\section{Conclusions}

We have presented high-precision polarization observations of $\alpha$~Oph at a range of wavelengths. We detect the wavelength dependent polarization expected for a rapidly-rotating star. We describe in detail the modelling procedure that allows us to predict the polarization expected for a rotating star and compare with the observations.

Our analysis of the A-type star $\alpha$~Oph does not constrain the parameters of the rotating star as well as was the case for our similar analysis of the B-type star Regulus \citep{cotton17}. This is because of the nature of the polarization wavelength dependence for a cooler star that does not include the distinctive reversal of sign that occurs in hotter stars. This makes it difficult to distinguish the effects of rotation rate from those of inclination. The analysis is further complicated by the need to fit the interstellar component of the polarization together with the rotating-star model. 

Nevertheless we are able to set significant constraints on the stellar properties. The rotation rate $\omega$ = $\Omega/\Omega_{\rm c}$ is found to be between 0.83 and 0.98 with an inclination between 60 and 90 degrees with best-fitting values following a correlation between the two as specified in Table~\ref{tab:tab_omeg}. These results are consistent with the interferometric determinations by \citet{zhao09}, but we find the rotation axis position angle to be 142$^\circ$ $\pm$ 4$^\circ$ different by 16$^\circ$ from the interferometric value. We suggest that the difference might arise from precession of the axis orientation due to interaction with the binary companion.

We also determine the polar gravity to be $\log{g_{\rm p}}$ = 3.93 $\pm$ 0.08. Comparison with rotating-star evolutionary models from \citet{georgy13} (see Fig.~\ref{fig:evol}) indicates that $\alpha$~Oph is in the later half of its main sequence evolution and must have been formed with an initial rotation rate of $\omega$ $\geq$ 0.8.

We determine the interstellar polarization towards $\alpha$~Oph to be characterized by a Serkowski law with $P_{\rm max}$ = 40.2 $\pm$ 7.1 ppm, $\theta$ = 21.0$^\circ$ $\pm$ 1.1$^\circ$ and $\lambda_{\rm max}$ = 440 $\pm$ 110 nm. This is one of the best determinations of interstellar polarization for such a nearby star and confirms earlier results showing a low value of $\lambda_{\rm max}$ for stars within the Local Hot Bubble \citep{cotton19b}.

\section*{Acknowledgements}

 We thank the AAT staff for support of observations at the Anglo-Australian Telescope. Funding for the construction of HIPPI-2 was provided by UNSW through the Science Faculty Research Grants Program. Based on observations under program GN-2018A-DD-108, obtained at the Gemini Observatory, which is operated by the Association of Universities for Research in Astronomy, Inc., under a cooperative agreement with the NSF on behalf of the Gemini partnership: the National Science Foundation (United States), the National Research Council (Canada), CONICYT (Chile), Ministerio de Ciencia, Tecnolog\'{i}a e Innovaci\'{o}n Productiva (Argentina), and Minist\'{e}rio da Ci\^{e}ncia, Tecnologia e Inova\c{c}\~{a}o (Brazil).




\bibliographystyle{mnras}
\bibliography{rasalhague}







\bsp	
\label{lastpage}
\end{document}